# Inference of chromosomal inversion dynamics from Pool-Seq data in natural and laboratory populations of *Drosophila melanogaster*

Martin Kapun[1*], Hester van Schalkwyk[1], Bryant McAllister[2], Thomas Flatt[1*] and Christian Schlötterer[1#]

[1]*Institut für Populationsgenetik, Vetmeduni Vienna, Veterinärplatz 1, A-1210 Vienna, Austria*

[*]*Present address: Department of Ecology and Evolution, University of Lausanne, Biophore, UNIL Sorge, CH-1015 Lausanne, Switzerland*

[2]*Department of Biology, University of Iowa, Iowa City, Iowa 52242, USA*

[#]*Corresponding author: Telephone: +43-1-25077-4300; Fax: +43-1-25077-4390; E-mail: christian.schloetterer@vetmeduni.ac.at*

Running title: Inversion frequencies in *Drosophila melanogaster* from Pool-Seq data

**Article summary**

Given their ability to suppress recombination, chromosomal inversions may be key factors shaping genetic variation. Therefore, they have been targeted by numerous studies and particularly regained attention since the advent of next generation sequencing. Here, we present a novel method to estimate inversion frequencies in Pool-Seq data of *D. melanogaster* based on inversion specific marker SNPs. We successfully applied this method to experimental evolution data and detected patterns consistent with positive selection. Moreover, we analyzed clinal frequency patterns along latitudinal gradients and found a previously unknown clinal pattern in a rare cosmopolitan inversion.

**Abstract**

**Sequencing of pools of individuals (Pool-Seq) represents a reliable and cost-effective approach for estimating genome-wide SNP and transposable element insertion frequencies. However, Pool-Seq does not provide direct information on haplotypes so that for example obtaining inversion frequencies has not been possible until now. Here, we have developed a new set of diagnostic marker SNPs for 7 cosmopolitan inversions in *Drosophila melanogaster* that can be used to infer inversion frequencies from Pool-Seq data. We applied our novel marker set to Pool-Seq data from an experimental evolution study and from North American and Australian latitudinal clines. In the experimental evolution data, we find evidence that positive selection has driven the frequencies of *In(3R)C* and *In(3R)Mo* to increase over time. In the clinal data, we confirm the existence of frequency clines for *In(2L)t*, *In(3L)P* and *In(3R)Payne* in both North America and Australia and detect a previously unknown latitudinal cline for *In(3R)Mo* in North America. The inversion markers developed here provide a versatile and robust tool for characterizing inversion frequencies and their dynamics in Pool-Seq data from diverse *D. melanogaster* populations.**

## Introduction

Inversions are common chromosomal variants in many organisms; they arise from structural mutations, which cause a complete reversal of linkage among genes relative to the standard chromosomal arrangement. Due to early efforts by Dobzhansky and his coworkers, much of our current understanding of the genetics and evolution of inversion polymorphisms comes from work on species of the genus *Drosophila* (Dobzhansky 1971; Powell 1997). Inversion polymorphisms are pervasive within numerous *Drosophila* species, and a large body of classical work suggests that they are important drivers of evolutionary dynamics and adaptive change in natural populations (for a review see Krimbas and Powell 1992).

Several lines of evidence indicate that selection plays a key role in maintaining inversion polymorphisms and in shaping their frequencies in natural populations. First, the frequencies of specific inversion polymorphisms in *Drosophila* have been correlated with numerous life history, physiological, and morphological traits (for reviews see Hoffmann *et al*. 2004, Hoffmann and Rieseberg 2008). Second, numerous polymorphic inversions show strongly clinal (e.g., latitudinal) patterns of variation, and many of these patterns are replicated across continents in broadly distributed species, including, *D. subobscura* (Prevosti *et al*. 1985; Prevosti *et al*. 1988; Krimbas and Powell 1992) *D. melanogaster* (Mettler *et al*. 1977; Knibb *et al*. 1981; Knibb 1982), and *D. pseudoobscura* (Dobzhansky and Epling 1944; Dobzhansky 1971; Powell 1997). Third, analyses of latitudinal gradients repeated over time indicate that many of these clines remain stable (Anderson *et al*. 1987) or that they shift with latitude over many years (Anderson *et al*. 2005; Levitan and Etges 2005). Finally, the fitness advantage and the dynamics of inversion heterokaryotypes have been monitored both in natural populations and under laboratory conditions, and the results are often consistent with selection shaping inversion dynamics (Wright and Dobzhansky 1946; Dobzhansky 1971). Moreover, inversions effectively suppress recombination around inverted regions in heterokaryotypes (Sturtevant 1917). Although double cross-overs and gene conversion can maintain a limited amount of gene flux between inverted and non-inverted arrangements (Chovnick 1973; Schaeffer and Anderson 2005), inversions typically cause a pattern of cryptic, chromosome-specific population substructure (Navarro et al. 2000). However, despite the large body of work on the population genetics of inversion polymorphisms, the nature of variation harbored by inversions and the molecular targets of selection within inversions remain very poorly understood to date (Hoffmann and Rieseberg 2008).

The recent advent of next-generation sequencing technology has revitalized interest in the population genetics of inversion polymorphisms. Several recent studies have used individual-based sequencing across multiple individuals to analyze the details of inversion breakpoint structure, the age of inversions, and patterns of genetic variation associated with inversions in natural populations (Corbett-Detig and Hartl 2012; Corbett-Detig et al. 2012; Langley et al. 2012). However, due to the still relatively high costs associated with sequencing many individuals, the availability of whole-genome population data for multiple individuals remains limited today. A widely used, very simple and cost-effective alternative is to sequence pools of DNA from multiple individuals ("Pool-Seq"; Futschik and Schlötterer 2010), but an obvious drawback of this approach is that it does not yield haplotype information and thus precludes the direct estimation of inversion frequencies.

Given the widespread use of the Pool-Seq method in molecular population genomics, and given the importance of inversions in shaping patterns of molecular variation in natural populations, here we have developed a novel set of SNP markers for 7 cosmopolitan inversions in *Drosophila melanogaster* (i.e. *In(2L)t*, *In(2R)Ns*, *In(3L)P, In(3R)C, In(3R)K, In(3R)Mo, In(3R)P*). By applying this new marker set to several natural and experimental populations we demonstrate that inversion frequencies and their dynamics can be reliably estimated from and examined with Pool-Seq data.

## Materials and Methods

We first developed a set of inversion-specific marker SNPs by karyotyping and whole-genome sequencing of individuals from an ongoing experimental evolution study in our laboratory (see Orozoco-terWengel *et al.* 2012; R. Tobler, V. Nolte, J. Hermisson, C. Schlötterer, unpublished results). To supplement this analysis, we also used haplotype information from the *Drosophila* Population Genomics Project (DPGP, DPGP2) (Langley *et al.*, 2012; Pool *et al.* 2012; http://www.dpgp.org; for details see below).

*Experimental evolution populations*

In brief, we carried out an experimental evolution experiment ("laboratory natural selection", LNS) using an outbred base population of *D. melanogaster* derived from 113 isofemale lines isolated from a wild population from Povoa de Varzim (Northern Portugal) in 2008 (see Orozoco-terWengel *et al.* 2012 for details; Tobler *et al.*, unpublished results). We exposed 3 replicate populations per treatment to two thermal selection regimes, with temperatures changing every 12 hours between 18°C and 28°C ("hot") and between 10°C and 20°C ("cold"). In both treatments, replicate populations were maintained with discrete generations at a fixed population size of 1000 individuals per replicate.

### Karyotyping

To determine the distribution of inversions in the above-mentioned selection experiment we used karyotyping. We randomly chose males of unknown chromosomal karyotype from three different cohorts: (1) isofemale lines, which were initially used to establish the base population of the experimental evolution experiment; (2) three replicate populations from the "cold" treatment at generation 34 of selection; and (3) three replicate populations from the "hot" treatment at generation 60 of selection. Males were crossed to virgin females of an inversion-free mutant strain (*y*[1]; *cn*[1] *bw*[1] *sp*[1]). In the F1, we prepared polytene chromosome squashes from salivary glands of 3$^{rd}$ instar larvae reared at 18°C using orcein staining following standard protocols (Kennison 2000). Chromosome preparations were analyzed using a Leica DM5500B microscope (Leica, Wetzlar, Germany). We determined chromosomal arms using reference maps in Bridges (1935) and Schaeffer *et al*. (2008); inversion loops in heterokaryons were identified from reference photographs in Ashburner *et al*. (1976). Corpses of some larvae used for chromosome preparations were stored in 96% EtOH for later DNA extraction and sequencing.

### Single individual sequencing

Based on information from our karyotyping, we selected 15 corpses of F1 larvae from three replicate populations of the hot and the cold selection regime at generations 60 and 34, respectively, for whole-genome sequencing (Supporting Table 1). We prepared individual genomic libraries by extracting DNA from homogenized single larva using the Qiagen DNeasy Blood and Tissue Kit (Qiagen, Hilden, Germany) and sheared DNA with a Covaris S2 device (Covaris Inc., Woburn, MA). To identify residual heterozygosity in the reference strain (*y*[1]; *cn*[1] *bw*[1] *sp*[1]) we sequenced a pool of 10 adult females. Each library was tagged with unique 8-mer DNA labels and pooled prior to preparation of a paired-end genomic library using the Paired-End DNA Sample Preparation Kit (Illumina, San Diego, CA); each library was sequenced on a HiSeq2000 sequencer (Illumina, San Diego, CA) (2 x 100 bp paired-end reads).

*Mapping of reads*

Raw reads were trimmed to remove low quality bases (minimum base quality: 18) using *PoPoolation* (Kofler *et al*. 2011) and mapped against the *D. melanogaster* reference genome (v.5.18) and *Wolbachia* (NC_002978.6) with *bwa* (v.0.5.7; Li and Durbin 2009) using the following parameters: –n 0.01 (error rate), -o 2 (gap opening), -d12, -e 12 (gap length) and -l 150 (disabling the seed option). We used the *bwa* module sampe to reinstate pair-end information using Smith-Waterman local alignment. Using *samtools* (Li *et al*. 2009), we merged SAM files filtered for proper pairs with a minimum mapping quality of 30 in a mpileup file and used *Repeatmasker* 3.2.9 (www.repeatmasker.org) to mask simple repetitive sequence and transposable elements (based on the annotation of the *D. melanogaster* genome v. 5.34). Using *PoPoolation*, we masked all indels (and five nucleotides flanking them on either side) present in at least one population and supported by at least two reads to avoid confounding effects of mis-mapping reads containing indels.

*Reconstitution of chromosomal haplotypes*

We used custom software tools to reconstruct paternal haplotypes from the sequenced F1 larvae (see above). By contrasting polymorphisms present in the F1 larvae to the reference sequence we inferred paternal alleles at heterozygous sites in F1 hybrids. Polymorphic positions (minimum minor allele frequency >10%) in reads from the reference strain (see above) were excluded. In addition, we used the following criteria to avoid false positive paternal alleles or false negative maternal alleles during haplotype reconstruction: (1) we excluded positions with a minimum coverage <15 to reduce false negatives due to large sampling error; (2) we calculated genome-wide coverage distributions for each F1 hybrid and each chromosomal arm separately and excluded positions with a coverage higher than the 95% percentile of the corresponding chromosomal arm to minimize false positives due to mapping errors and duplications; (3) we only included alleles with a minimum count of 20 across all larvae sequenced; (4) for SNPs with more than two alleles we only considered the two most frequent alleles; (5) we only retained alleles for which the allele counts fell within the limits of a 90% binomial confidence interval based on an expected frequency of 50%. The efficiency of our SNP calling was evaluated using two different methods (see Supporting Information).

*Fixed differences associated with inversions*

We took advantage of a worldwide sample of haplotypes originating from Africa, Europe and North America with known karyotype (Langley *et al*. 2012; Pool *et al*. 2012) and combined them with our haplotype data. In total, we compared 167 chromosomes from Africa (DPGP2; 107 individuals), Portugal (present study; 15 individuals), France (DPGP2; 8 individuals) and USA (DPGP; 37 individuals [consensus genomes]) with known karyotypes, overall representing 7 different inversions (*In(2L)t*, *In(2R)Ns*, *In(3L)P*, *In(3R)C*, *In(3R)K*, *In(3R)Mo*, *In(3R)P*) plus standard chromosome arrangements (Supporting Table 2). For each inversion type, we searched for fixed differences in the combined dataset between inverted karyotypes and all other arrangements (i.e., standard arrangements and overlapping inversions) on the corresponding chromosome in order to identify inversion-specific SNP markers. We excluded positions where less than 80% of all individuals per arrangement were informative. We tested our method as described in the Supporting Information.

*Inversion frequency estimates*

We used inversion-specific fixed differences between arrangements as SNP markers to estimate inversion frequencies from Pool-Seq datasets of Fabian *et al*. (2012; North American cline), Kolaczkowski *et al*. (2011; Australian cline), Orozco-terWengel *et al*. (2012; experimental evolution experiment, "hot" selection regime) and Tobler *et al*. (unpublished results; experimental evolution, "cold" regime). Inversion frequencies were estimated from the average of all marker allele frequencies specific to a particular inversion. To reduce the variance in frequency estimates caused by sampling error we excluded all positions with less than 10-fold coverage for all datasets except for the Australia data, where – given the generally low coverage in this dataset – we chose a minimum coverage threshold of three-fold. We also excluded all positions with coverage larger than the 95% percentile of the genome-wide coverage distribution to avoid errors due to mis-mapping or duplications. To evaluate the statistical significance of inversion frequency differentiation over time in our experimental evolution study, we integrated SNP-wise allele frequency information from three replicate populations in each selection regime across multiple timepoints by performing Cochran-Mantel-Haenzel tests (CMH; Landis *et al*. 1978) for each marker SNP separately and by averaging *P*-values across all tests. Since replicates were not available for the two latitudinal datasets, we performed Fisher's Exact Tests (FET; Fisher 1922) on inversion frequency differences between the

249  lowest-latitude population and all other populations along each cline (North America,
250  Australia) and combined *P*-values across all marker SNPs. We also compared
251  inversion frequency estimates obtained from SNP markers to our empirical results
252  from karyotyping as described in the Supporting Information. In addition, we also
253  estimated inversion frequencies from our karyotype data and tested for significant
254  differences in inversion frequency between the "hot" and "cold" selection regimes by
255  using the following fully factorial fixed-effects two-way ANOVA model: $y = I + T +$
256  $I \times T$, where *y* denotes the inversion frequency, *I* the inversion type and *T* the
257  selection regime using JMP (v.10.0.0, SAS Institute Inc., Cary, NC).

258

259  *Genetic variation within inversions*
260  To estimate genetic variation associated with each chromosomal arrangement we
261  estimated $\pi$ in 100-kb non-overlapping sliding windows for all chromosomes with the
262  same karyotype. We excluded *In(2R)Ns* and *In(3R)P* from this analysis since both
263  inversions were present in only one F1 larva out of the 15 sequenced individuals. To
264  compare $\pi$ among arrangements we randomly sub-sampled non-inverted chromosomes
265  to match the number of inverted chromosomes for *In(2L)t* and *In(3L)P*. For the
266  inversions on *3R* (*In(3R)Mo* and *In(3R)C*) we were unable to sub-sample because our
267  dataset only contained three chromosomes with standard arrangement on this
268  chromosomal arm. We therefore used all three individual chromosomes to estimate $\pi$
269  and $F_{ST}$ among chromosomal arrangements on *3R*. In addition, based on our estimates
270  of $\pi$, we calculated $F_{ST}$ between inverted and standard arrangement haplotypes in 100-
271  kb non-overlapping windows to measure the amount of chromosome-wide
272  differentiation among arrangements.

273

*Linkage disequilibrium within inversions*

For each chromosomal arm and arrangement, we estimated linkage disequilibrium (LD) by calculating $r^2$ (Hill and Robertson 1968). We randomly sampled 5000 polymorphic SNPs along each chromosomal arm and visualized chromosome-wide pair-wise $r^2$-values using heat maps generated from the 'LDHeatmap' package (Shin *et al*. 2006) in *R* (R Development Core Team 2009). To quantify the difference in overall LD within non-inverted and inverted chromosomes, we averaged all $r^2$-values obtained from within the inverted regions for both standard and inverted haplotypes separately and calculated their ratios. Since $r^2$ depends strongly on the number of haplotypes, we always matched the number of inverted and standard chromosomes by sub-sampling the more frequent chromosomal arrangement.

*Expected inversion frequency change under neutrality*

To estimate the degree to which inversion frequency changes observed during experimental evolution may be explained by drift alone, we employed forward simulations using a simple Wright-Fisher model of neutral evolution (Otto and Day 2007). For computations, we considered an inversion to represent allele *A*. Inversion frequencies $p_0(A)$ at the beginning of the experiment were obtained from frequency estimates based on our marker SNP approach. Additionally, we used estimates of the effective population size computed from real data of the laboratory natural selection experiment and performed simulations using a value of 200 for the parameter *N* (Orozco-terWengel *et al*. 2012). Using 100,000 iterations we simulated all three replicate populations for each temperature regime and using the same number of generations and the inversion frequency from the base population. We computed the empirical p-value by determining the number of simulations in which the polarized frequency change in each of the replicates was larger than in the observed data.

## Results

### *Impact of inversions on genetic variation*

In total, we identified six polymorphic cosmopolitan inversions segregating in our experimental evolution experiment: four common inversions (*In(2L)t*, *In(2R)Ns*, *In(3L)P*, *In(3R)Payne*) and two rare cosmopolitan inversions (*In(3R)Mo*, *In(3R)C*) (see Mettler *et al*. 1977; Lemeunier and Aulard 1992). We first aimed to examine the partitioning of genetic variation among inversions and standard chromosomes by performing whole-genome sequencing of 15 out of 275 karyotyped individuals and by reconstructing the paternal haplotypes of these flies (see Materials and Methods; Supporting Table 1). We estimated nucleotide diversity (□) and LD ($r^2$) for both inverted and non-inverted chromosomes and calculated pairwise $F_{ST}$ to estimate genetic differentiation between arrangements. Since *In(2R)Ns* and *In(3R)P* were only represented by one chromosome in our data, we did not analyze these inversions.

   *2L*: $\pi$ was similar between the standard arrangement and *In(2L)t* except for the breakpoint regions, where inverted chromosomes were less variable than the standard arrangement. $F_{ST}$ was markedly higher within the inversion breakpoints as compared to the outside of the inverted region (see Supporting Figure 1a), but did not show distinct peaks at the putative breakpoints. Pairwise $r^2$ values along *2L* indicated the existence of elevated LD in two regions located within the inversion and at the telomeric end of the chromosomal arm in haplotypes carrying *In(2L)t*. LD within inverted haplotypes was 2.46 times higher within the chromosomal region of the inversion as compared to standard arrangement chromosomes (see Supporting Figure 2a).

*3L*: In contrast to standard arrangement chromosomes, we found reduced variability ($\pi$) around the proximal breakpoint of *In(3L)P* and in two large regions within the inversion as well as downstream of the distal breakpoints in chromosomes carrying the inverted arrangement. Although $F_{ST}$ was homogenous along the chromosome, we detected an unusual haplotype structure in the *In(3L)P* chromosomes, with very large areas of pronounced LD within the inversion and also extending beyond it (see Supporting Figure 1b and Supporting Figure 2b). Overall, LD within inverted haplotypes was approximately 4.7 times higher than in standard chromosomes.

*3R*: We found four chromosomal arrangements on the right arm of the third chromosome segregating in the populations from the selection experiment (standard arrangement, *In(3R)C*, *In(3R)Mo*, *In(3R)Payne*, all of which are known to overlap; Lemeunier and Aulard 1992). In contrast to chromosomes carrying *In(3R)C* and *In(3R)Mo*, the standard arrangement chromosomes did not exhibit any regions of reduced heterozygosity (Figure 1). *In(3R)Mo* karyotypes harbored almost no genetic variation within the inverted region, except for two polymorphic regions with a size of approximately 1 and 2 mb, respectively (see Supporting Information for details). Moreover, 2 mb upstream of the proximal breakpoint the *In(3R)Mo* karyotypes were almost completely genetically invariant. We also observed a large haplotype ranging from more than 6 mb upstream to approximately 1 mb downstream of *In(3R)Mo*. In contrast to *In(3R)Mo*, *In(3R)C* did not show any continuous genomic regions exhibiting highly reduced genetic variation. Nonetheless, genetic variation was locally reduced at the breakpoints of the two overlapping inversions *In(3R)Mo* and *In(3R)Payne*. The strongest reduction, showing almost complete absence of genetic variation, was found in a region of approximately 500 kb close to the distal breakpoint of *In(3R)Mo*. However, apart from locally elevated haplotype structure at the proximal breakpoint of *In(3R)C* and the telomeric part of *3R*, we did not observe elevated levels of LD (see Figure 2B). Pairwise $F_{ST}$ was increased for both inverted karyotypes within the inversions as well as in their proximity. Interestingly, we observed peaks of differentiation only at the proximal breakpoints of both inversions but not at their distal breakpoints. Moreover, despite pronounced haplotype structure in *In(3R)Mo*, we failed to find strong differences in LD between the different arrangements (*In(3R)Mo*, LD ratio: 1.05; *In(3R)C*, LD ratio: 1.13).

*Identification of inversion-specific SNPs*

Next, we used our data to define inversion-specific SNPs that could be used as diagnostic markers for detecting and surveying specific inversions. Alleles private to *In(2L)t*, *In(3L)P*, *In(3R)K*, and *In(3R)Payne* were almost entirely restricted to the inversion breakpoints (Figure 3). In contrast, alleles specific to *In(2R)Ns* and *In(3R)C* were distributed throughout these inversions (Figure 3). For *In(3R)Mo*, we not only found marker SNPs within the inversion but also a surplus of SNPs beyond the proximal and distal breakpoints (Figure 3). The number of marker SNPs in the different inversions varied greatly, ranging from 4 in *In(3R)K* to 150 in *In(3R)Mo* (Supporting Table 3). Importantly, two complementary methods for detecting false positives and a comparison of inversion frequency estimates based on karyotyping versus marker SNPs indicated that our SNP marker set is highly reliable (Supporting Information).

372 *Inversion dynamics during experimental evolution*

373 We used these inversion-specific marker SNPs to investigate the dynamics of

374 inversions during our experimental evolution experiment, using three replicate

375 populations in each selection regime. For each inversion, we estimated its frequency

376 by averaging over the frequencies of all inversion-specific SNP markers. With a

377 baseline frequency of about 40% in the base population, *In(2L)t* was the most

378 frequent inversion in the experiment. Its frequency fluctuated unpredictably across

379 selection regimes and replicate populations, but the inversion remained polymorphic

380 throughout the experiment with frequencies larger than 20% (see Figure 4, Supporting

381 Figure 3A, Supporting Table 4). In contrast, *In(2R)Ns* started out at a frequency of

382 approximately 10% in the base populations and then consistently decreased in all

383 replicates in both selection regimes (Figure 4, Supporting Figure 3B, Supporting

384 Table 4). This pattern resulted in a statistically significant difference in inversion

385 frequency between the base population and the third time point examined in both

386 thermal selection regimes (Supporting Table 5). Similarly, *In(3R)Payne* decreased

387 significantly in frequency in both regimes (see Figure 4, Supporting Figure 3G,

388 Supporting Table 4), a trend already noticed by Orozco-terWengel *et al*. (2012) for

389 the "hot" regime. Interestingly, three inversions showed a selection regime-specific

390 behavior. While *In(3L)P* remained stable around 15% in the "cold" regime, it

391 decreased significantly over time in the "hot" regime (Figure 4, Supporting Figure

392 3C, Supporting Table 4). In contrast *In(3R)Mo* initially segregated at a very low

393 frequency of approximately 5% in the base populations but then consistently

394 increased to >25% in all replicates of the "cold" regime while showing inconsistent

395 frequency patterns in the "hot" regime (Figure 4, Supporting Figure 3F). Finally,

396 *In(3R)C* started out at approximately 15%, then strongly increased over time in all

397    replicates of the "hot" regime, but fluctuated unpredictably in the "cold" regime

398    (Figure 4, Supporting Figure 3D). In good agreement with these changes in inversion

399    frequencies as estimated from our SNP markers, we found highly significant effects

400    of inversion type (2-way ANOVA, $F_{5,24} = 21.339$, $P < 0.0001$) and of the inversion

401    type by selection regime interaction ($F_{5,24} = 6.9793$, $P < 0.001$) in our data based on

402    inversion frequencies observed from 275 karyotyped larvae, confirming again the

403    reliability of our novel inversion-specific SNP markers.

404

405    *Spatial distribution of inversions in natural populations*

406    We next used our inversion-specific SNPs to estimate inversion frequencies in two

407    previously published Pool-Seq datasets of populations collected along latitudinal

408    clines in North America (Fabian *et al*. 2012) and Australia (Kolaczkowski *et al*.

409    2011). For the North American data we found a clinal distribution of most inversions

410    (Supporting Figure 4A, Supporting Table 6). *In(2L)t*, *In(3L)P* and *In(3R)Payne*

411    showed strongly clinal patterns negatively correlated with latitude (Supporting Table

412    7). While *In(2L)t* and *In(3L)P* decreased linearly from south (Florida) to north

413    (Maine), *In(3R)Payne* was very frequent (~50%) in Florida but almost absent in

414    Pennsylvania and Maine (also see Fabian *et al*. 2012). In contrast, the frequencies of

415    *In(2R)Ns*, *In(3R)K* and *In(3R)Mo* increased with latitude. *In(3R)C* segregated at very

416    low frequencies and showed no clinal pattern.

417         Similarly, we estimated inversion frequencies for the two endpoints of the parallel

418    but independent Australian cline (Queensland and Tasmania; cf. Kolaczkowski *et al*.

419    2011) (Supporting Figure 4B, Supporting Table 6). Similar to the patterns we

420    observed for the North American cline, we found that *In(2L)t*, *In(3L)P* and

421    *In(3R)Payne* were much more frequent at low latitude (Queensland) but absent or at low frequency at high latitude (Tasmania). However, none of the observed frequency differences were significant according to FET (see Supporting Table 7), maybe due to the low sequence coverage in this dataset. We did not detect the presence of *In(2R)Ns*, *In(3R)C*, *In(3R)K* and *In(3R)Mo* in the Australian dataset, but due to low coverage we were unable to determine whether these inversions occur at a very low frequency or whether they are truly absent.

**Discussion**

Numerous previous studies have aimed to understand patterns of genetic variation associated with inversions in *D. melanogaster* (e.g., see Andolfatto *et al*. 2001, and references therein). Fixed genetic differences associated with inversions have been of particular interest since they may provide valuable information about the evolutionary history of these structural variants. For example, variation around inversion breakpoints has frequently been used to estimate inversion age (e.g., Hasson and Eanes 1996; Andolfatto *et al*. 1999; Matzkin *et al*. 2005). However, previous studies have been limited by the restricted amount of available data, and especially by the paucity of reliable molecular markers for detecting and surveying inversions in *D. melanogaster*.

    Here, we have aimed to extend these efforts by using a combination of next-generation whole-genome sequence analysis and classical karyotyping of inversions in *D. melanogaster*. Specifically, by combining haplotype data from our present study (based on both individual-level sequencing and karyotyping) with publicly available haplotype information from known karyotypes in the DPGP and DPGP2 data, we have developed a new and extensive set of inversion-specific marker SNPs. These novel diagnostic markers have allowed us characterize the frequency dynamics of 7

polymorphic inversions in both laboratory and natural populations of *D. melanogaster*.

*Patterns of divergence in chromosomal inversions*

Overall, we observed large heterogeneity in the number and distribution of divergent SNPs for different inversions. In three of the common large cosmopolitan inversions (*In(2L)t*, *In(3L)P* and *In(3R)Payne*) and in the rare large cosmopolitan inversion *In(3R)K* we found only few divergent SNPs, most of which were restricted to the inversion breakpoints. These patterns agree well with previous observations for *In(2L)t* and *In(3L)P* (Andolfatto *et al*. 2001) and provide further evidence that suppression of gene flux is mainly restricted to only a few kb around the inversion breakpoints.

    For *In(2R)Ns*, which is also considered to be common cosmopolitan inversion and which has a similar age as *In(2L)t*, *In(3L)P*, *In(3R)Payne* and *In(3R)K* (Corbett-Detig and Hartl 2012), we identified fixed differences throughout the whole inversion. This inversion is markedly smaller than the other cosmopolitan inversions (~4.8 mb), resulting in an effective recombination rate of approximately 18cM across the inverted region (e.g., Fiston-Lavier *et al*. 2010; Comeron *et al*. 2012). Since double crossing-over is unlikely to occur in regions of less than 20cM (Navarro *et al*. 1997), presumably because the minimum distance between chiasmata is limited by crossing-over interference (McPeek and Speed 1995), the pattern we have observed for *In(2R)Ns* might reflect low rates of gene conversion.

    Similar to *In(2R)Ns*, we found that for two rare cosmopolitan inversions on *3R* (*In(3R)C*, *In(3R)Mo*) fixed differences were also not restricted to the breakpoint regions. *In(3R)C* is a large terminal inversion (> 12mb), and marker SNPs for this

inversion showed a pronounced non-homogeneous distribution. SNPs were found across the distal half of the inverted region, perhaps reflecting reduced recombination close to the telomere rather than an inversion-specific pattern. Alternatively, this pattern might reflect selection of co-adapted *In(3R)C*-specific alleles. However, since *In(3R)C* haplotypes were only available from one population from Portugal, we cannot rule out that these patterns are highly specific.

The number and distribution of marker SNPs for *In(3R)Mo* differed markedly from all other inversions. For this inversion, we detected the highest number of marker SNPs and found them to be distributed inside the inversion as well as beyond the inversions boundaries, both proximal and distal. This strongly suggests that suppression of recombination occurs well beyond the inversion breakpoints.

*Distribution of inversions in natural populations*

The pervasive clinal distribution of the cosmopolitan inversions *In(2L)t*, *In(3L)P* and *In(3R)Payne* along latitudinal gradients is well-known and has been documented for numerous populations in North America, Australia, and Asia already over 30 years ago (Knibb 1982). The fact that qualitatively similar frequency clines for these inversions have been observed on multiple continents has been taken as strong *prima facie* evidence for the non-neutral maintenance of these inversions by spatially varying selection. However, up-to-date no conclusive data have been published about whether the clinal patterns for these inversions have remained stable or not. While two studies from Australia (Anderson *et al*. 1987; Anderson *et al*. 2005) found that inversion clines remained stable or shifted with latitude, a study from Japan observed pronounced changes in some populations over many years (Inoue *et al*. 1984). We were therefore interested in using our inversion-specific SNP markers to examine inversion frequencies in recently generated Pool-Seq data for the North American (Fabian *et al*. 2012) and Australian (Kolaczkowski *et al*. 2011) clines.

Despite a large difference in sequence coverage between these two recent studies (approximately 45-fold versus 11-fold coverage), we observed clinal frequency patterns for *In(2L)t*, *In(3L)P* and *In(3R)Payne* that are in excellent qualitative agreement with previous findings from the 1970s and 1980s (Mettler et al. 1977; Knibb et al. 1981; Knibb 1982) for both the Australian and the North American cline. Remarkably, our data suggest that the inversion frequencies for *In(3R)Payne* and *In(3L)P* have remained extremely stable for more than 30 years. In contrast, for *In(2L)t* we also observed clinal variation but detected an increase in the frequency of this inversion by approximately 20% in all populations as compared to previous observations. Although we observed strong inversion clines in the data from the Australian east coast that are qualitatively consistent with previous studies, our inversion frequency estimates for Australia were generally lower than those reported in previous work. While it is possible that these results reflect a reduction in inversion frequencies in Australia in recent years, we cannot rule out that the low sequencing coverage of the Australian data has downward-biased our estimates. Clearly, further in-depth analysis of these inversions will be necessary to understand the mechanisms that determine their dynamics and maintenance.

*In(2R)Ns*, in contrast, showed a different pattern to that observed for *In(2L)t*, *In(3L)P* and *In(3R)Payne*. Two earlier studies found this inversion to occur at a frequency of >20% in Queensland (Mettler et al. 1977; Knibb et al. 1981), but our analysis of the Australian data suggests that this inversion has either decreased to very low frequencies or that it has completely vanished in Australia. For the North American cline we also found a pattern that contrasts with previous results: Mettler *et* *al*. (1977) reported that the frequency of *In(2R)Ns* decreases with increasing latitude, whereas in our analysis this inversion showed a weakly (non-significant) clinal trend from approximately 0-1% frequency in Florida up to 7-10% in Maine.

The three rare cosmopolitan inversions *In(3R)C*, *In(3R)K* and *In(3R)Mo*)were either not present in the Australian data or segregated at frequencies below our detection threshold. In contrast, for the North American east coast, we found both *In(3R)C* and *In(3R)K* to be segregating at very low frequencies, consistent with previous observations (Mettler *et al*. 1977; Knibb 1982). Surprisingly, while *In(3R)Mo* was found to be very rare and non-clinal in North America 30 years ago (e.g., Mettler *et al*. 1977) we now detect a positive correlation with latitude. This is consistent with the data of Langley *et al*. (2012) who have recently noticed a considerable increase in *In(3R)Mo* frequency (up to a frequency of approximately 18% in Raleigh, North Carolina). Together, our data indicate that *In(3R)Mo* has recently undergone a strong increase in frequency along the North American east coast. Although the reasons for this striking pattern remain unclear, the strong reduction of genetic variation within and around *In(3R)Mo* described here and in two other recent studies (Langley *et al*. 2012; Corbett-Detig and Hartl 2012) is consistent with this notion and indicates a recent origin coupled with a rapid increase in frequency.

***Implications of inversion polymorphisms for genome scans of selection***

Our investigation of inversion frequency dynamics during experimental evolution clearly demonstrates that the frequencies of some inversions change consistently among replicate populations. While some inversions decreased in frequency in both thermal selection regimes, three of them changed consistently in frequency in only one of the selection regimes. A meta-analysis of inversion frequency changes during experimental evolution by Inoue (1979) has reported that inversion frequencies generally decrease during experimental evolution. However, in contrast to Inoue (1979), we found two inversions (*In(3R)C* and *In(3R)Mo*) whose frequencies clearly and consistently increased over time in one of the selection regimes in our experimental evolution study. Wright-Fisher simulations of neutral evolution based on the initial inversion frequencies show that frequency changes observed for these two inversions were significantly higher than expected due to genetic drift alone (see Supporting Table 8). Thus, this pattern strongly suggests that both inversions must likely have carried one or several selection regime-specific favorable alleles. Interestingly, and perhaps consistent with a selective role for this inversion, *In(3R)C* has previously been shown to affect bristle number variation in an artificial selection experiment (Izquierdo *et al*. 1991), yet we did not monitor this phenotype in our experimental evolution study.

    In a genome-wide analysis of our "hot" selection regime, Orozco-terWengel *et al*. (2012) have identified the majority of candidate SNPs to be located on chromosome *3R*, which also harbors four overlapping inversions. Strikingly, two of these inversions, *In(3R)C* and *In(3R)Payne*, changed significantly in frequency in the "hot" regime over the experiment, perhaps suggesting that beneficial alleles in these inversions have been major targets of selection. Yet, among the most significant candidate SNPs identified by Orozco-terWengel *et al*. (2012) only 1-3 of the marker SNPs for *In(3R)C* (depending on the dataset analyzed) overlapped the candidate SNPs sets. If the inversion was the only cause for the strong molecular signature of selection on *3R* in this experiment, these inversion-specific SNPs would clearly be expected to show the largest allele frequency differences, yet they do not. Instead, we hypothesize

that the presence of inversions in laboratory populations can result in cryptic chromosome-specific population structure which in turn causes elevated drift and leads to a surplus of candidate SNPs. If selection is assumed to operate on top of this structure, the interpretation of the SNP data becomes very challenging. Thus, even though the inversions might play an important role in the response to selection, distinguishing the effects due to selection from those due to population structure is practically difficult. One way around this problem in experimental evolution studies using *Drosophila* would be to use inversion-free *Drosophila* species.

In natural populations we have observed a similar phenomenon. Despite almost all sites being shared between *In(3R)Payne* and the non-inverted chromosome, populations with a high *In(3R)Payne* frequency seem to harbor more variation (also see Fabian *et al.* 2012), as might be expected for a subdivided population. Since inverted and non-inverted chromosomes will have different allele frequencies, the contrast of populations with different inversion frequencies for the inference of selection is also challenging. On the other hand, in our previous study of clinal variation along North American cline, we found 77% of all clinal candidate SNPs to be located on *3R* and >50% of the candidates within the region spanned by *In(3R)Payne*, a highly non-random pattern that is consistent with spatially varying selection (Fabian *et al.* 2012) and that is also qualitatively mirrored in the Australian data (Kolaczkowski *et al.* 2011). Nonetheless, due to the difficulty of teasing apart the effects of demography and population structure versus those of selection, we anticipate that in the future genome scans of selection might preferentially focus on chromosomes with the same inversion status or use inversion-free systems.

## Conclusions

Here we have presented a novel and robust set of molecular SNP markers for seven polymorphic chromosomal inversions in *D. melanogaster*, which will be highly useful for the analysis of Pool-Seq data in this model. Although overall we have found a good correlation between our SNP-based and karyotype-based inversion frequency estimates, we would like to caution that our inference of inversion-specific SNPs is highly dependent on the available reference genomes. In particular, for *In(3R)C*, *In(3R)K* and *In(3R)Mo*, which did not occur in all populations in our combined dataset, we cannot rule out that our marker SNP sets contain some false positives. Therefore, for diverged populations, inversion frequency estimates may be less accurate. Yet, given that multiple SNPs contribute to the estimates of inversion frequencies, we expect that our set of inversion-specific markers will show a reliable performance across all *Drosophila* populations.

## Data accessibility

The raw FASTQ files for all single individuals are available from the European Sequence Read Archive (http://www.ebi.ac.uk/ena/about/search_and_browse) under the accession number XXX.

Reconstructed haploid genomes (available as FASTA files) as well as a collection of custom written Python scripts are available from the Dryad database (http://datadryad.org) under the accession: XXX

622   **Acknowledgements**

623   We thank all the members of the Institute of Population Genetics for their support, in
624   particular Andrea Betancourt for helpful discussion, Raymond Tobler for sharing
625   unpublished data, and Daria Martynow, Daniel Fabian and Raymond Tobler for help
626   with karyotyping. Above all, we are grateful to Viola Nolte for library construction
627   and handling the sequencing data. Our work was supported by the Austrian Science
628   Foundation (FWF P19467 grant to CS) and the Swiss National Science Foundation
629   (SNF PP00P3_133641 grant to TF). BFM was supported by a Fulbright grant from
630   the Austrian-American Educational Commission.

631

**Figure Legends**

**Figure 1. Nucleotide diversity (π) and genetic differentiation ($F_{ST}$) for *In(3R)Mo* and *In(3R)C*.** Line plots showing nucleotide diversity (π) in standard (blue) and inverted (red) chromosomal arrangements; additionally, $F_{ST}$ values (black) show the amount of genetic differentiation between arrangements. (A) *In(3R)Mo* (based on five individuals). (B) *In(3R)C* (based on six individuals). Values for standard arrangement chromosomes (blue) were obtained from comparing three individual chromosomes. Putative boundaries of the three overlapping inversions on *3R* are indicated by vertical black lines: the dashed line represents *In(3R)Mo*, the dotted line *In(3R)P* and the solid line *In(3R)C*.

**Figure 2. Linkage disequilibrium for *In(3R)Mo* and *In(3R)C*.** Triangular heatmaps showing estimates of $r^2$ for 5000 randomly sampled SNPs across *3R*. The bottom triangles show the results for inverted arrangements, whereas the top triangles show the standard arrangements (based on three individuals). (A) $r^2$ plots for *In(3R)Mo* (based on 5 individuals). (B) $r^2$ plots plots for *In(3R)C* (based on 6 individuals). The chromosomal position of the three overlapping inversions on *3R* is indicated by a colored line: *In(3R)P* (red), *In(3R)Mo* (blue), and *In(3R)C* (black).

**Figure 3. Distribution of fixed SNPs within inversions.** Chromosomal distribution of inversion-specific differences based on a global sample of 167 haplotypes . The number of divergent SNPs is binned in 100-kb non-overlapping sliding windows and plotted along the chromosomal arm carrying the corresponding inversion. Vertical dashed lines indicate the putative inversion breakpoints.

**Figure 4. Inversion frequency trajectories during experimental evolution.**

Inversion frequencies estimated by marker SNPs from Pool-Seq data for the three different replicate populations in each selection regimes ("cold" indicated by dashed and "hot" indicated by solid lines) of our laboratory natural selection experiment. The frequency estimates were calculated by averaging the frequencies of all marker allele for each inversion separately.

804 **Table 1. Inversion counts and frequencies.** Counts and frequencies (in parentheses) of six inversions identified by karyotyping in the base

805 population and three replicate populations in each selection regime. The sample size *n* refers to the number of chromosomes sampled from each

806 population.

| Population | n | *In(2L)t* | *In(2R)Ns* | *In(3L)P* | *In(3R)P* | *In(3R)Mo* | *In(3R)C* |
|---|---|---|---|---|---|---|---|
| Base | 37 | 12 (0.32) | 2 (0.05) | 1 (0.03) | 4 (0.11) | 4 (0.11) | 5 (0.14) |
| cold - R1 | 36 | 13 (0.36) | 0 (0) | 3 (0.08) | 3 (0.08) | 7 (0.19) | 2 (0.06) |
| cold - R2 | 45 | 4 (0.09) | 0 (0) | 2 (0.04) | 0 (0) | 12 (0.27) | 12 (0.27) |
| cold - R3 | 30 | 10 (0.33) | 2 (0.07) | 0 (0) | 0 (0) | 6 (0.2) | 3 (0.1) |
| hot - R1 | 42 | 15 (0.36) | 0 (0) | 2 (0.05) | 0 (0) | 2 (0.05) | 19 (0.45) |
| hot - R2 | 44 | 10 (0.23) | 0 (0) | 3 (0.07) | 2 (0.05) | 1 (0.02) | 15 (0.34) |
| hot - R3 | 41 | 16 (0.39) | 0 (0) | 0 (0) | 0 (0) | 1 (0.02) | 17 (0.41) |
| Sum | 275 | 80 | 4 | 11 | 9 | 33 | 73 |

807

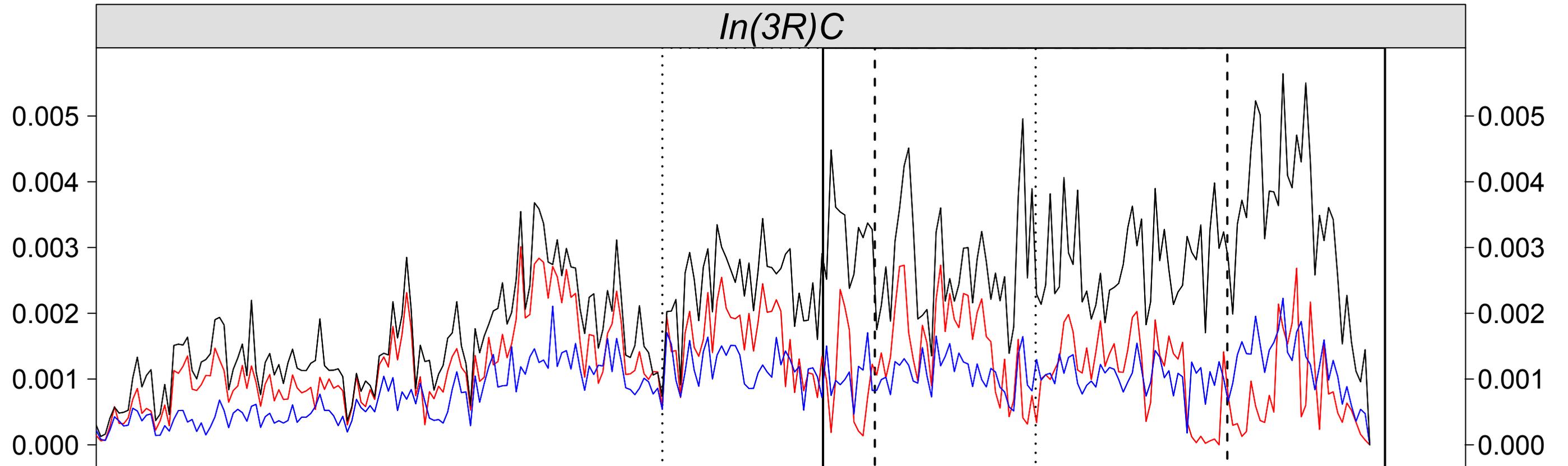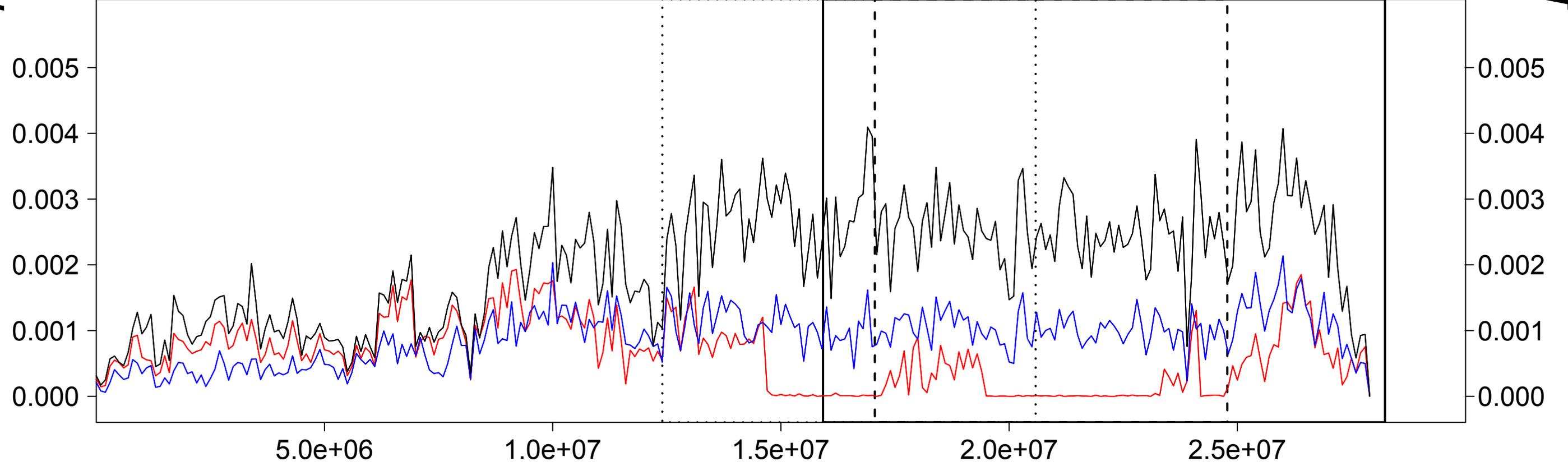

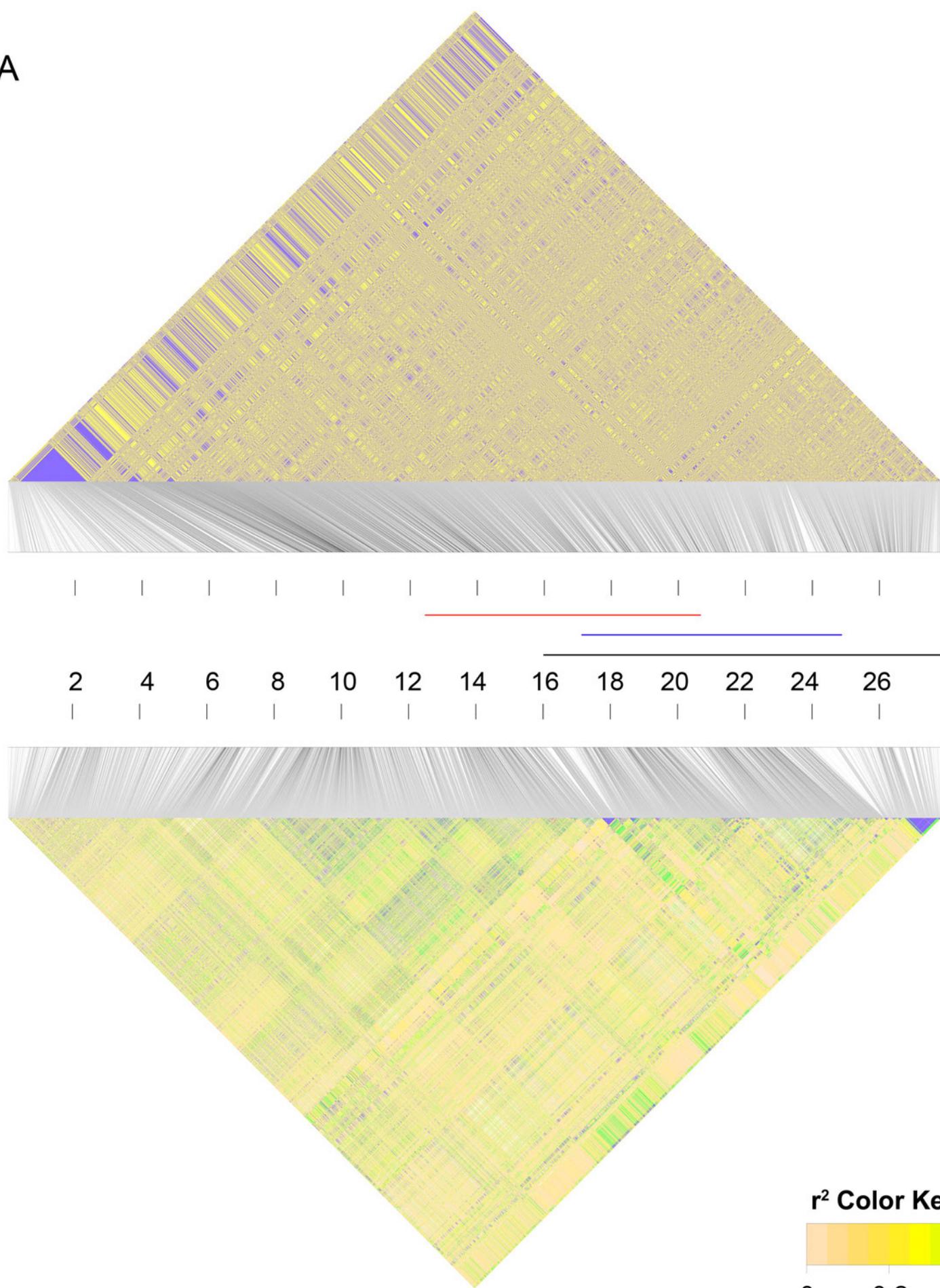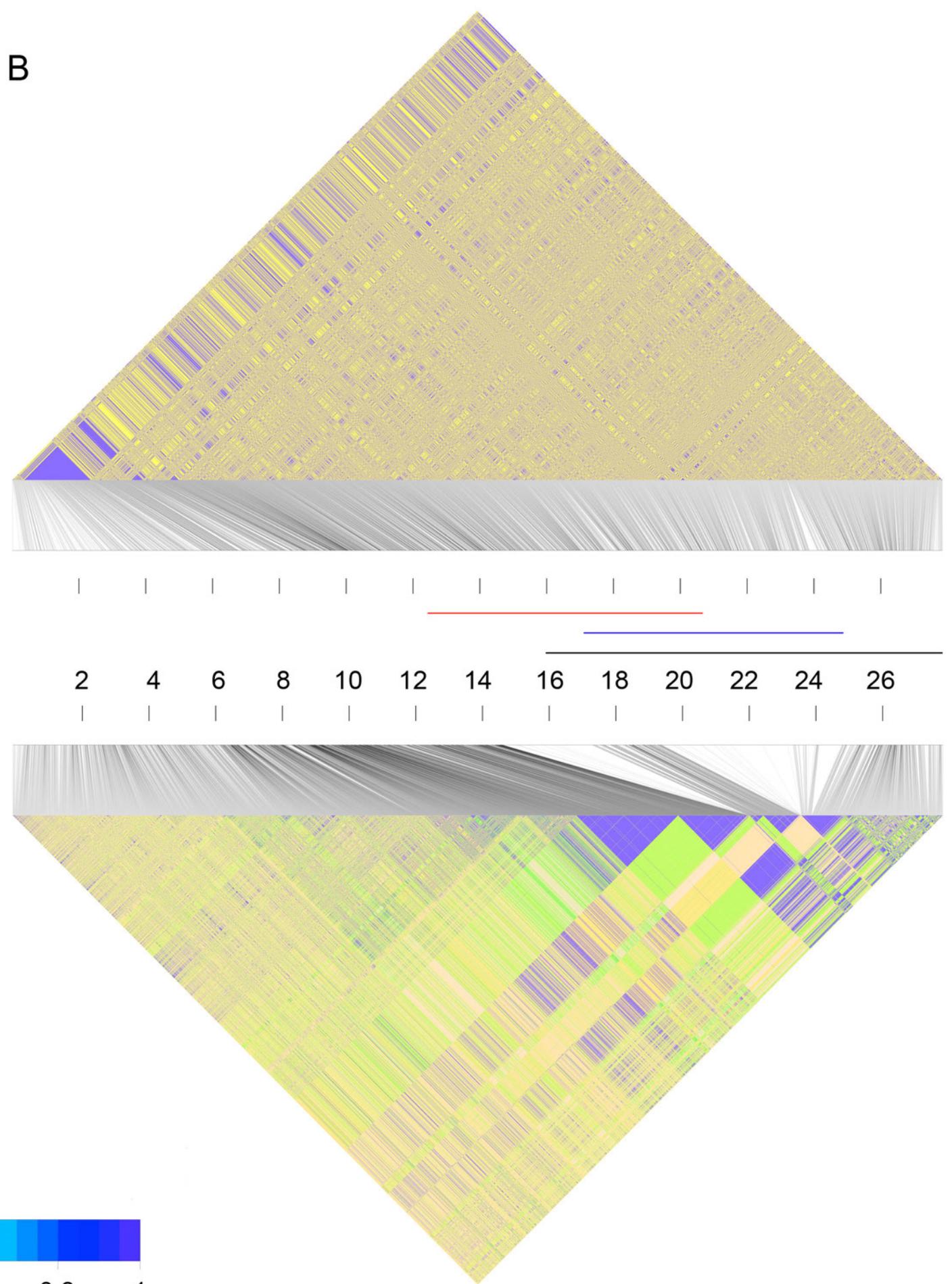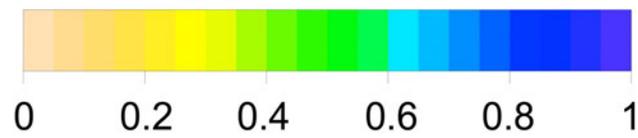

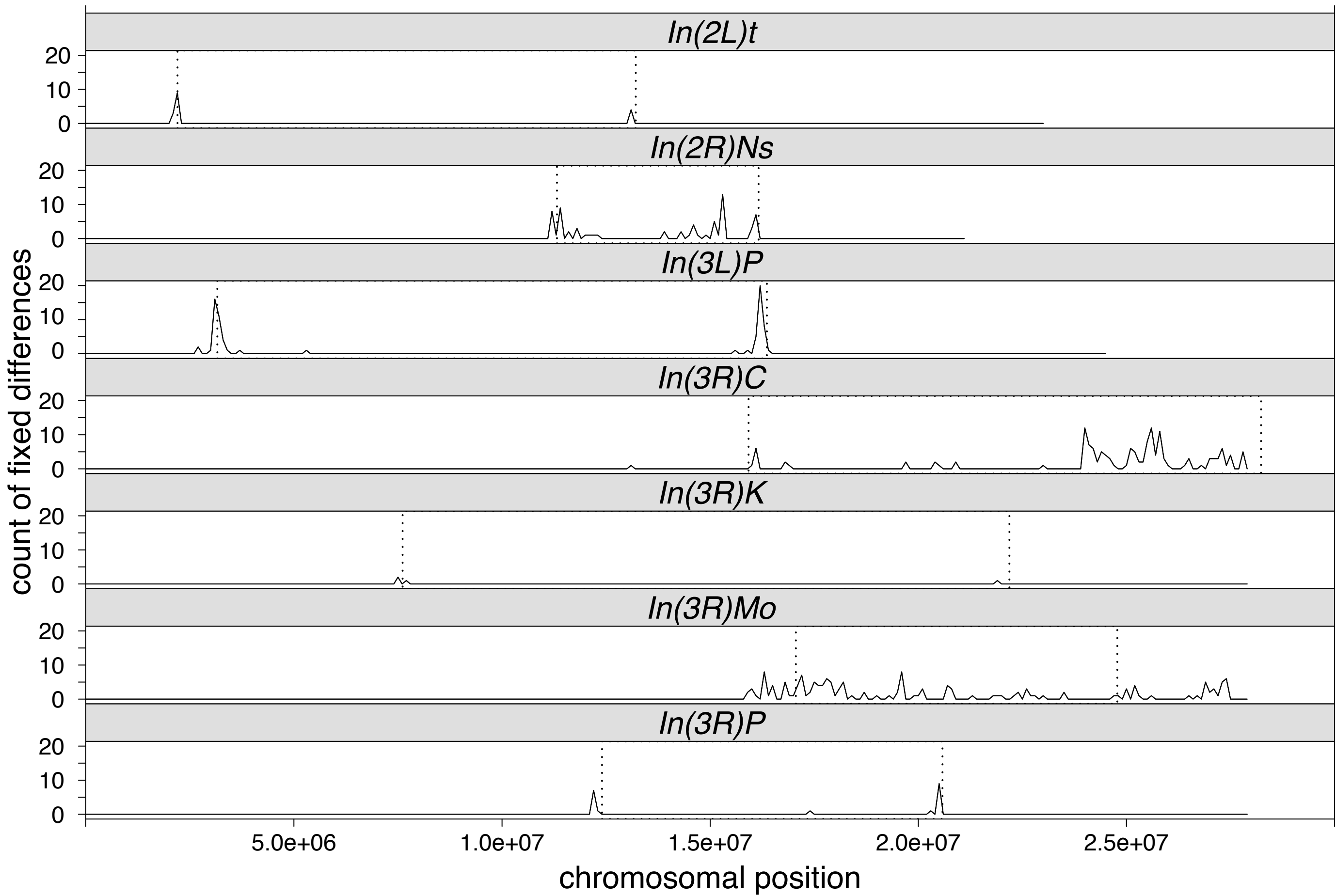

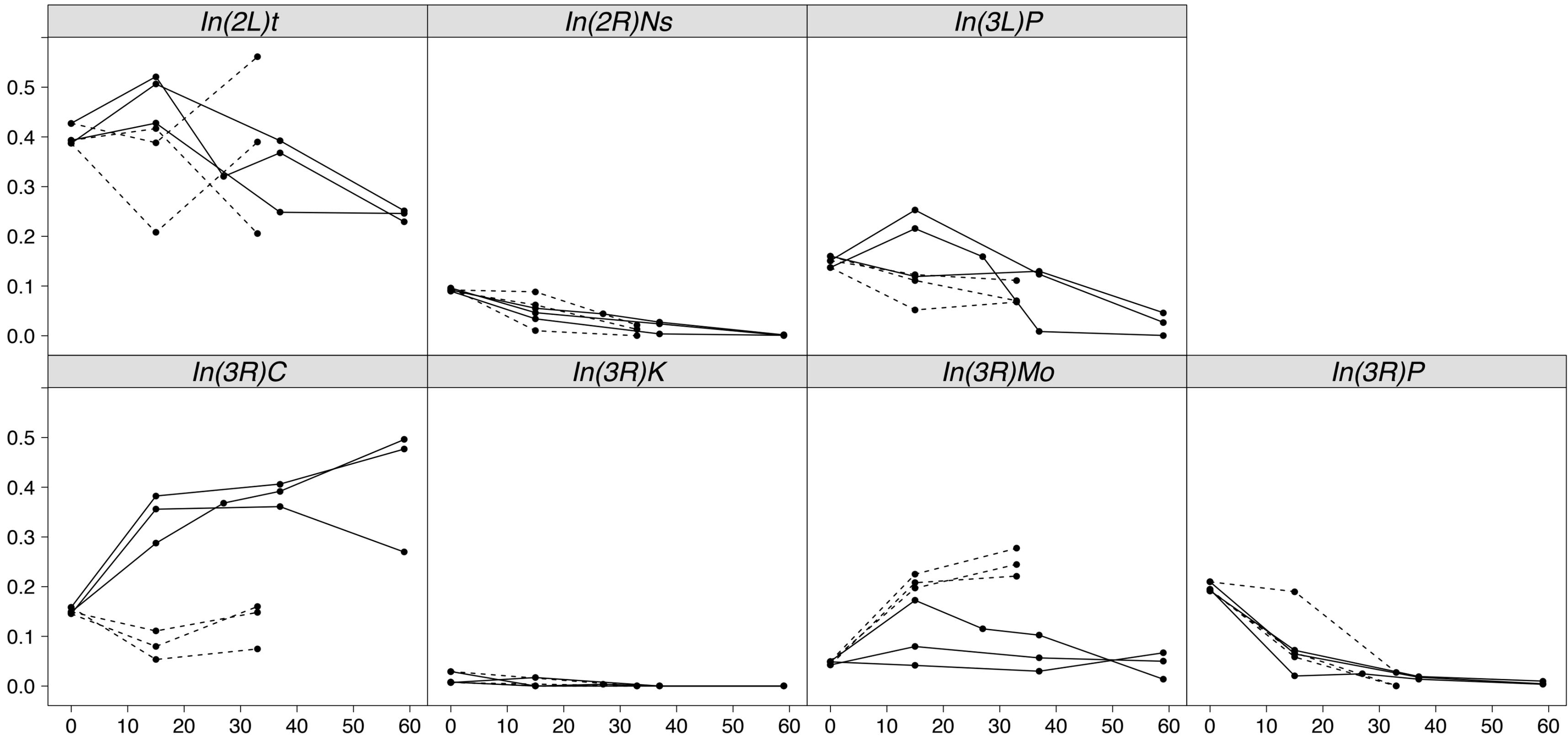

# 1 **Supporting Information**

**Estimation of false negative and false positives during haplotype reconstruction**

19  Based on our crossing scheme for chromosomal karyotyping, we developed a novel
20  bioinformatics pipeline to reconstruct sire (male parent) haplotypes from whole-
21  genome-sequenced F1 larvae. As described in the Material and Methods section, we
22  implemented several filtering and stringency thresholds to avoid wrongly typed
23  alleles. Here we describe two methods, which were used to estimate the number of
24  false positives and false negatives among reconstructed haplotypes. First, we sexed
25  sequenced larvae based on cytology and sequencing data: male *Drosophila*
26  individuals are homozygous for the *X* chromosome, which results in (i) large DNA
27  staining intensity differences between autosomes and the *X* in preparations of
28  polytene chromosomes and (ii) large coverage differences between autosomes and the
29  *X* in next-generation sequencing data. With these two methods, we were able to
30  unambiguously identify two male larvae in our dataset. In these individuals, only the
31  maternal copy of the *X* chromosome was sequenced; thus, all SNPs detected on the *X*
32  in these individuals represent sequencing or mapping errors. These data therefore
33  allowed us to estimate the overall false positive rate. For individual number 136
34  (approximately 48-fold autosomal coverage) and individual number 100
35  (approximately 27-fold autosomal coverage) we detected 9 and 13 false positive SNPs
36  respectively, translating into false positive rates of $4 \times 10^{-7}$ and $5 \times 10^{-7}$ along the *X*
37  chromosome (approximately 22.4 mb long) for the parameter combinations used in
38  the analysis. Supporting Figure 5 shows the false positive rate for four different
39  parameter combinations for both male individuals. Second, in single individuals
40  sequenced with next-generation sequencing allele frequencies of polymorphic SNPs
41  are distributed around a frequency of 0.5 depending on sequencing depth. However,

42   low coverages inflate the sampling error, which can result in the absence of

43   polymorphic alleles. Given that we sequenced the reference strain used for the

44   crosses, we were able to identify cases among the F1 hybrid sequences for which

45   positions appeared to be fixed for an allele different than the reference. Assuming that

46   the distribution of frequencies caused by sampling error is symmetrical, we were able

47   to obtain false negative rates for our data. Supporting Figure 6 shows the average

48   coverages and false negative rates for each individual at different minimum coverage

49   thresholds. In summary, our results strongly suggest that the haplotype datasets used

50   in our analysis were not affected by high false positive and false negative rates.

52   **Number of false positives in inversion-specific fixed differences**

53   In our study we developed a panel of inversion-specific fixed SNP markers, obtained

54   by analyzing karyotype-specific nucleotide variation in an alignment of 167 *D.*

55   *melanogaster* genomes originating from Africa, Europe and North America (see

56   Supporting Table 1). To rule out false positives due to sampling artifacts, we

57   estimated false positive rates using permutations. We randomly assigned individuals

58   as being inverted or non-inverted a 100 times (in the same proportions as in the real

59   data) and counted the number of falsely identified candidates. None of the permuted

60   data resulted in any false positive candidate SNPs.

61       We further tested whether the inversion-specific markers SNPs identified inversion

62   frequency differences more accurately than randomly selected SNPs located within

63   the boundaries of corresponding inversions. We therefore performed CMH tests

64   between the base population and consecutive experimental generations in both

65   selection regimes for each marker SNP separately, as described in Materials and

66   Methods. To obtain a combined result we averaged over all $\chi^2$ values. We then

67  randomly sampled 10,000 times the same number of SNPs as the real marker SNPs
68  and performed CMH tests; for each of these 10,000 sets we counted how often the $\chi^2$
69  values from the random data were larger than for the marker SNPs. By sampling from
70  the tails of this distribution we obtained empirical *P*-value estimates, based on a cut-
71  off defined by the $\chi^2$ value of the real marker SNPs. Under the null hypothesis,
72  inversion-specific alleles would be expected to not perform better in predicting
73  inversion frequencies than randomly drawn samples from within the inversion. The
74  empirical *P*-values from this analysis are shown in Supporting Table 11. We found
75  that our marker SNPs performed significantly better than randomly drawn SNPs for
76  those inversions whose frequencies changed most strongly over time in our selection
77  experiment (i.e., *In(3R)P* and *In(2R)Ns* in both regimes; *In(3R)Mo* in the "cold"
78  regime; and *In(3R)C* in the "hot" regime), but not for inversions whose frequencies
79  changed only weakly or which were segregating at very low baseline frequencies.
80
81  **Reliability of using inversion-specific fixed differences as inversion-specific**
82  **markers in Pool-Seq data**
83  Next, we examined the extent to which our fixed marker SNPs provide accurate
84  estimates of inversion frequencies in our Pool-Seq data. To do so, we compared
85  empirical data based on karyotyping of flies from our laboratory natural selection
86  experiment with inversion frequencies estimated from our Pool-Seq data. Using
87  Fisher's exact tests (FET) we asked whether inversion frequency counts obtained
88  from karyotyping differ significantly from the average inversion frequency counts as
89  estimated by our inversion-specific SNP markers. None of the 36 tests (6 inversions
90  × 2 treatments × 3 replicates; Supporting Table 9) resulted in *P*-values <0.05.
91  Therefore, our results clearly suggest that our set of inversion-specific marker SNPs is

92  very reliable and robust in terms of accurately estimating inversion frequencies from
93  Pool-Seq datasets.
94
95  **Complex patterns of gene flux and genetic variation in overlapping inversions**
96  The presence of three overlapping inversions on *3R* in our haplotype data provides a
97  unique opportunity for studying genetic exchange between different arrangements.
98  We focused on *In(3R)Mo* which was represented by 5 chromosomes in our dataset.
99  With the exception of two polymorphic regions within the inversion boundaries,
100 *In(3R)Mo* showed almost complete absence of genetic variation within and beyond
101 the inversion boundaries (see Figure 1). We identified two individuals (numbers 96
102 and 100) which carried polymorphisms within the inversion body of *In(3R)Mo* (see
103 Supporting Figure 7A). To further explore the genealogical relationship among all
104 chromosomes with different arrangements in these two polymorphic regions, we
105 reconstructed phylogenetic trees based on $\pi$, using only SNPs with unique alleles in
106 individuals 96 and/or 100 (see Supporting Figure 7A-C). Therefore, we constructed
107 distance matrices by calculating average $\pi$ for all possible chromosome pairs in the
108 sample and used the neighbor-joining method to generate dendrograms using the *R*
109 package 'ape' (Paradis *et al*. 2004). We determined the statistical significance of each
110 node by bootstraping 1000 times, each time randomly drawing a subset corresponding
111 to 10% of all SNPs from the dataset, and then calculated consensus trees using 'ape'
112 in *R*.
113 Interestingly, in all phylogenies either one or both of these individuals differed
114 significantly from all other *In(3R)Mo* chromosomes. Specifically, in the proximal half
115 of the first polymorphic region, both individuals were highly similar and clustered
116 with the standard arrangement and with the single *In(3R)Payne* individual (see

117   Supporting Figure 7A), whereas individual 100 only clustered with the chromosome
118   carrying *In(3R)Payne* in the distal half (see Supporting Figure 7B). In contrast, in the
119   second region only individual 96 clustered with standard arrangement chromosomes
120   (see Supporting Figure 7C). To further analyze the amount of allele sharing between
121   the different arrangements, we extracted SNPs specific to both individuals and
122   counted how often these alleles segregated in other arrangements. Remarkably, the
123   alleles specific to individual 96 were entirely shared with the standard arrangement
124   but not associated with a single haplotype. Similarly, the majority of alleles (>75 %)
125   specific to individual 100 from the first region were also shared with the standard
126   arrangement. A major proportion of the alleles specific to both individuals was also
127   shared with *In(3R)C* and with the single individual carrying *In(3R)Payne* (see
128   Supporting Table 10). In summary, these findings indicate that the patterns observed
129   within *In(3R)Mo* haplotypes are the result of multiple recent recombination events, at
130   first between different arrangements and subsequently between *In(3R)Mo* haplotypes.
131

135
136
137
138
139
140
141

Kapun *et al.,* Supporting Information File

**Supporting Figures and Tables**

**Supporting Figure 1. Nucleotide diversiy ($\pi$) and genetic differentiation ($F_{ST}$) for *In(2L)t* and *In(3L)P*.** Line plots showing $\pi$ averaged in 100-kb non-overlapping sliding windows of individuals with standard (blue) and inverted (red) chromosomal arrangement; $F_{ST}$ values (black) show the amount of genetic differentiation between these arrangements. (A) results for *In(2L)t*, for five individuals of each karyotype. (B) results for *In(3L)P*, for six individuals of each karyotype. In both (A) and (B), the black boxes represent the putative boundaries of the corresponding inversions.

**Supporting Figure 2. Linkage disequilibrium for *In(2L)t* and *In(3L)P*.** Triangular heatmaps showing the values of pairwise calculations of $r^2$ for 5000 randomly sampled SNPs across each chromosome. The bottom half shows the results for individuals with the inverted arrangement, whereas the top half shows the results for standard arrangement chromosomes, based on the same number of individuals as for the inverted karyotype. The chromosomal location of each inversion is highlighted as a red line. (A) Plots for *2L*, with *In(2L)t* at the bottom and the standard arrangement at the top (based on 5 individuals). (B) Plots for *3L*, with *In(3L)P* at the bottom and the standard arrangement at the top (based on 4 individuals).

**Supporting Figure 3. Inversion frequency trajectories during experimental evolution.** Box plots showing the allele frequency distributions of inversion-specific SNP markers across different selection regimes (rows; "hot" and "cold") and replicate populations (columns) in our laboratory natural selection experiment. We used the median of each distribution to estimate inversion frequencies. (A) Results for *In(2L)t*;

Kapun *et al.,* Supporting Information File

**Supporting Figures and Tables**

**Supporting Figure 1. Nucleotide diversiy ($\pi$) and genetic differentiation ($F_{ST}$) for *In(2L)t* and *In(3L)P*.** Line plots showing $\pi$ averaged in 100-kb non-overlapping sliding windows of individuals with standard (blue) and inverted (red) chromosomal arrangement; $F_{ST}$ values (black) show the amount of genetic differentiation between these arrangements. (A) results for *In(2L)t*, for five individuals of each karyotype. (B) results for *In(3L)P*, for six individuals of each karyotype. In both (A) and (B), the black boxes represent the putative boundaries of the corresponding inversions.

**Supporting Figure 2. Linkage disequilibrium for *In(2L)t* and *In(3L)P*.** Triangular heatmaps showing the values of pairwise calculations of $r^2$ for 5000 randomly sampled SNPs across each chromosome. The bottom half shows the results for individuals with the inverted arrangement, whereas the top half shows the results for standard arrangement chromosomes, based on the same number of individuals as for the inverted karyotype. The chromosomal location of each inversion is highlighted as a red line. (A) Plots for *2L*, with *In(2L)t* at the bottom and the standard arrangement at the top (based on 5 individuals). (B) Plots for *3L*, with *In(3L)P* at the bottom and the standard arrangement at the top (based on 4 individuals).

**Supporting Figure 3. Inversion frequency trajectories during experimental evolution.** Box plots showing the allele frequency distributions of inversion-specific SNP markers across different selection regimes (rows; "hot" and "cold") and replicate populations (columns) in our laboratory natural selection experiment. We used the median of each distribution to estimate inversion frequencies. (A) Results for *In(2L)t*;

167   (B) for *In(2R)Ns*; (C) for *In(3L)P*; (D) for *In(3R)C*; (E) for *In(3R)K*; (F) for *In(3R)Mo*

168   and (G) for *In(3R)Payne.* We performed CMH tests to test for significant frequency

169   differences between generation 0 and consecutive generations in the experimental

170   evolution experiment for each candidate SNP separately. Combined results were

171   obtained by averaging across all *P*-values of all marker SNPs. Green stars indicate

172   significant results between the base population (generation 0) and the corresponding

173   evolved populations at subsequent timepoints during the selection experiment (*

174   $P<0.05$, ** $P<0.01$, *** $P<0.001$).

175

176   **Supporting Figure 4. Inversion frequencies in natural populations.** Box plots

177   showing allele frequencies of inversion specific SNP markers in latitudinal

178   populations from Australia (A; Kolaczkowski *et al*. 2011) and North America (B;

179   Fabian *et al.* 2012). We performed Fisher's Exact tests (FET) to test for significant

180   frequency differences between the population at the lowest latitude (i.e., Florida and

181   Queensland, respectively) and all other populations along each cline for each

182   candidate SNP separately. Combined results were obtained by averaging across all *P*-

183   values of all marker SNPs. Green stars indicate significant results for the comparison

184   between the lowest-latitude population and the other populations (* $P<0.05$, **

185   $P<0.01$, *** $P<0.001$).

186

187   **Supporting Figure 5. False positive rates in haplotype reconstruction.** False

188   positive rates estimated for two male F1 hybrids (individuals 100 and 136) for

189   different filtering parameters (minimum allele count and minimum mapping quality),

190   as described in Materials and Methods; also see Supporting Text for further details.

191

192  **Supporting Figure 6. False negative rates in haplotype reconstruction.** Average
193  coverages based on next-generation sequencing data for the reference strain and all 15
194  F1 hybrids (grey line) and false negative rate estimates for different minimum
195  coverage thresholds for each individual separately. See Supporting Text for further
196  details.
197
198  **Supporting Figure 7. Patterns of recombination within *In(3R)Mo*.** The center plot
199  shows $\pi$ averaged in 100-kb non-overlapping sliding windows for three different
200  combinations of individuals carrying *In(3R)Mo* within the inverted region on *3R*. The
201  orange line represents individuals 80, 129 and 150; the black line the three former
202  individuals plus individual 100; and the grey line individuals 80,129, 150 and 96.
203  Dendrograms were generated from distance matrices based on $\pi$ calculated for all
204  pairwise comparisons using SNPs with unique alleles in individuals 96 or 100. The
205  chromosomal arrangements of individuals in the trees are color-coded, with *In(3R)Mo*
206  shown in red, *In(3R)C* in green, *In(3R)Payne* in blue and the standard arrangement in
207  black. We used bootstrapping to test for the consistency of the tree topologies.
208  Branches with >95% bootstrapping support are indicated with a purple dot. Trees in
209  (A) and (B) are based on SNPs specific for individual 96, whereas (C) is based on
210  SNPs with unique alleles in individual 100. The length of the scale bar in each plot
211  corresponds to $\pi = 0.1$.
212
213

214

## Supporting Figure 1

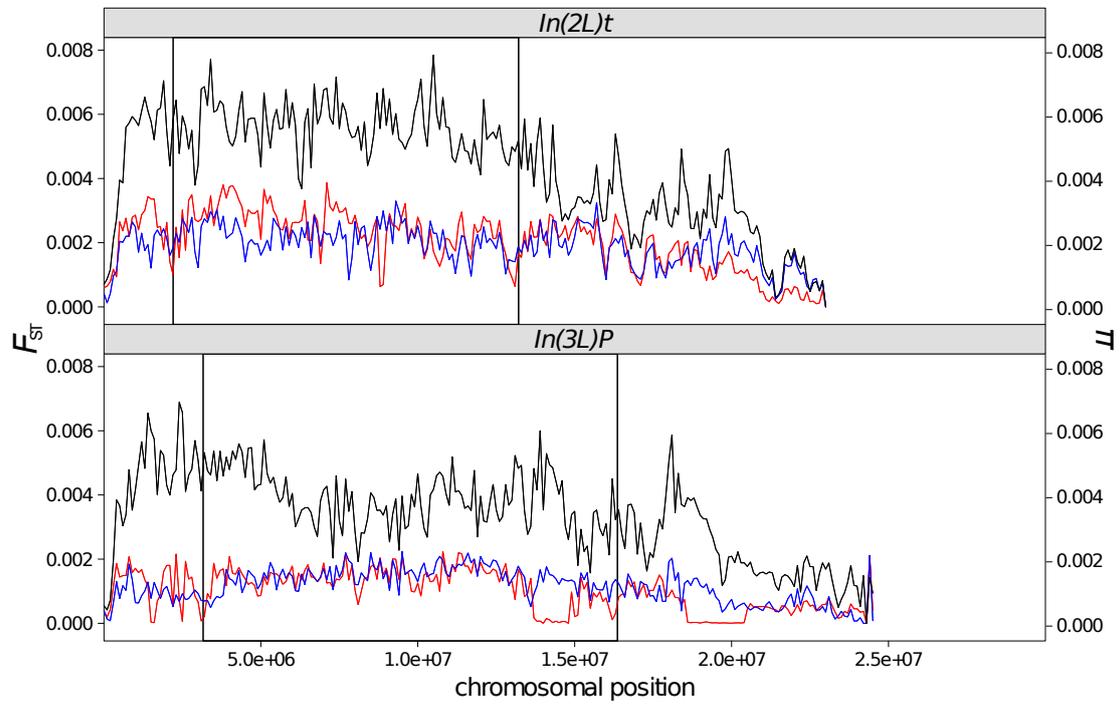

216

217 **Supporting Figure 2**

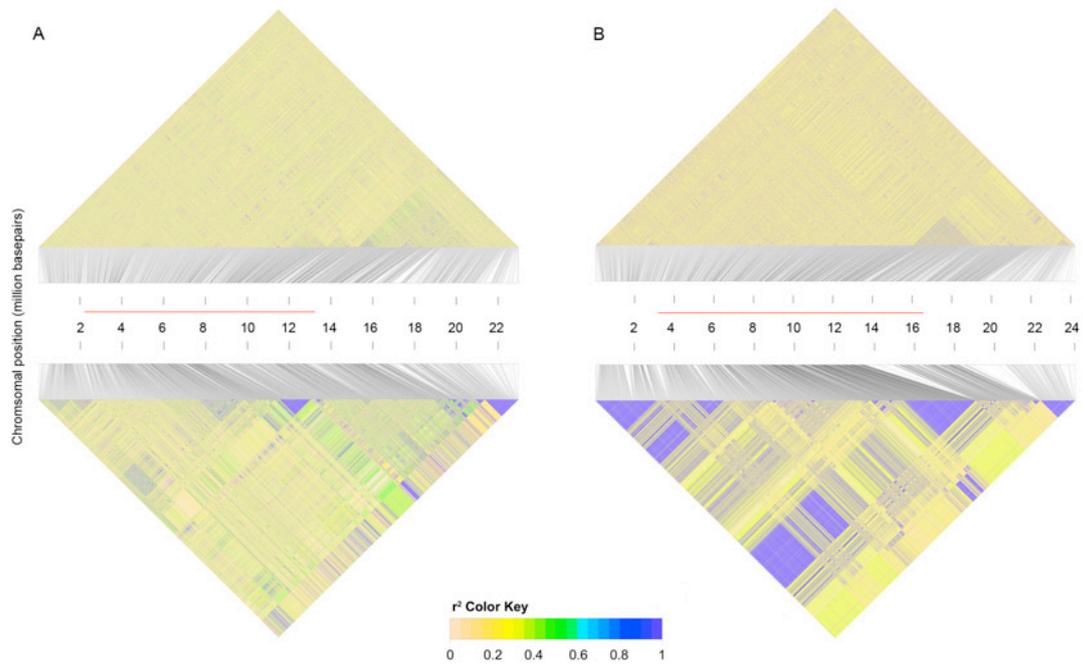

218

**Supporting Figure 3**

**A**

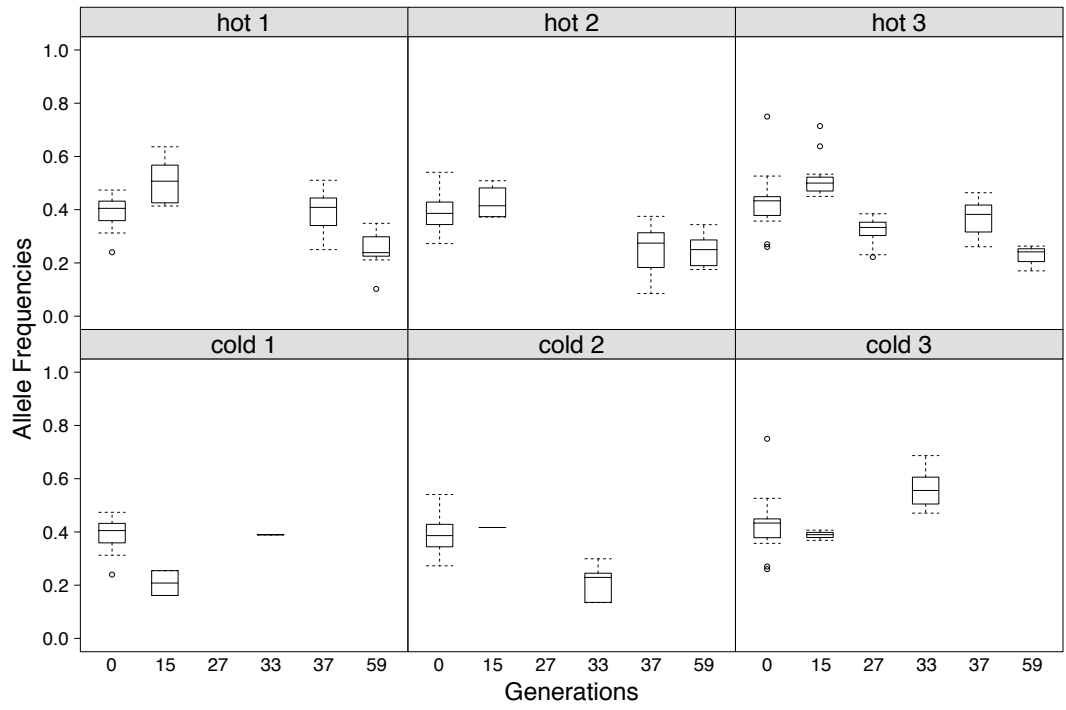

**B**

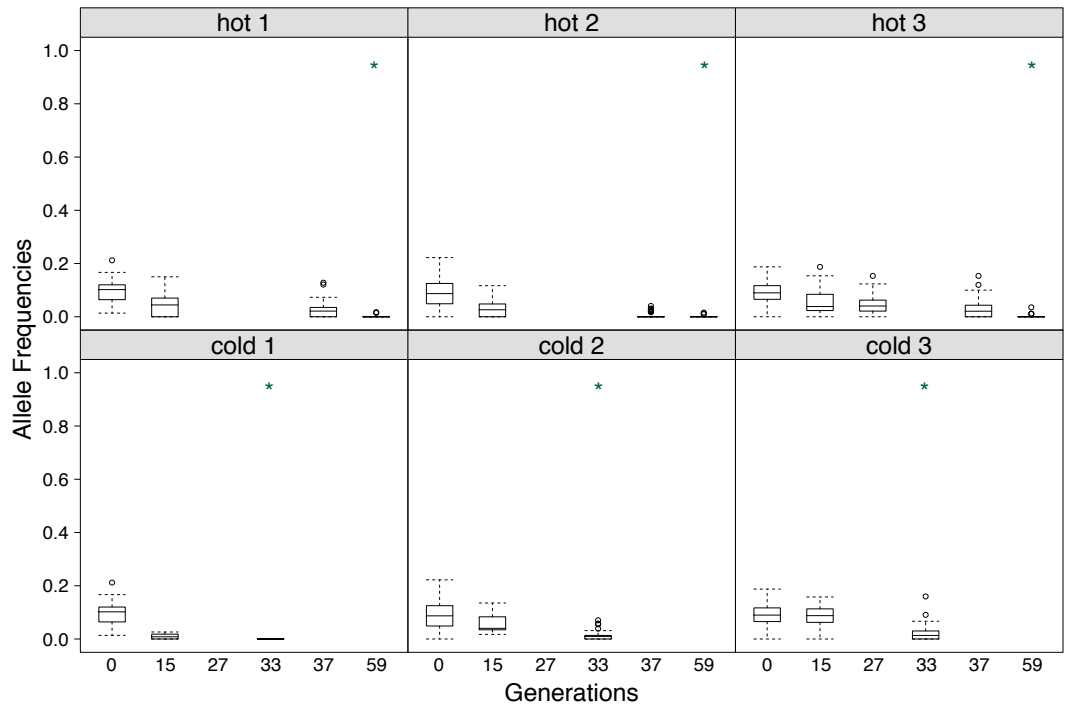

225 **C**

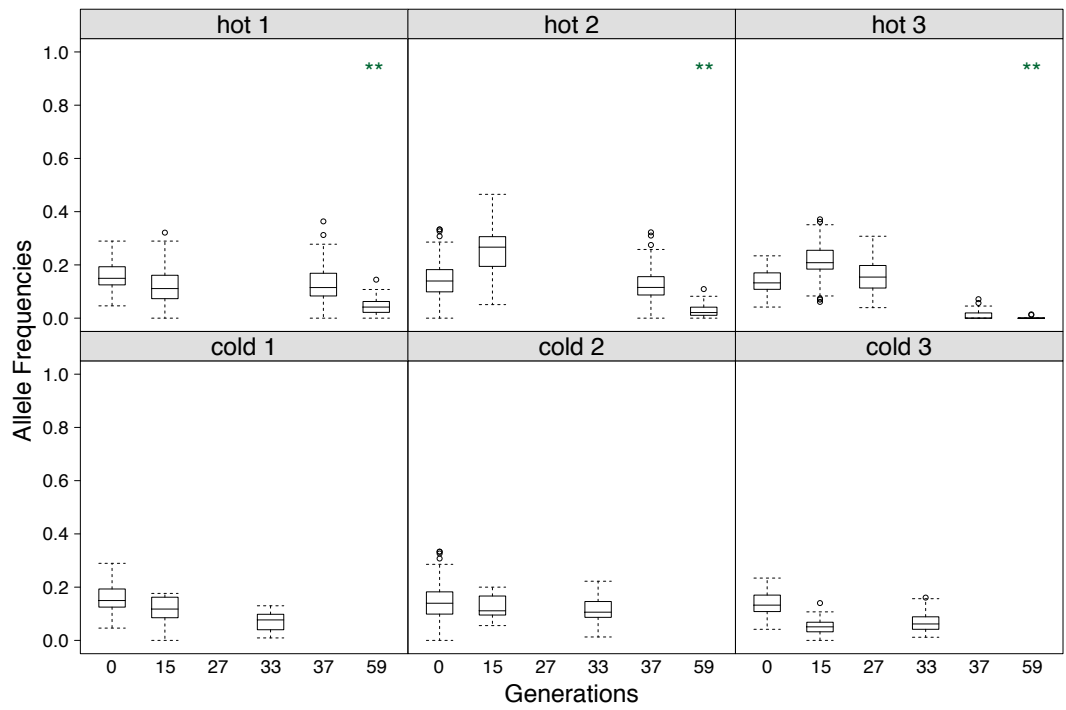

226

227 **D**

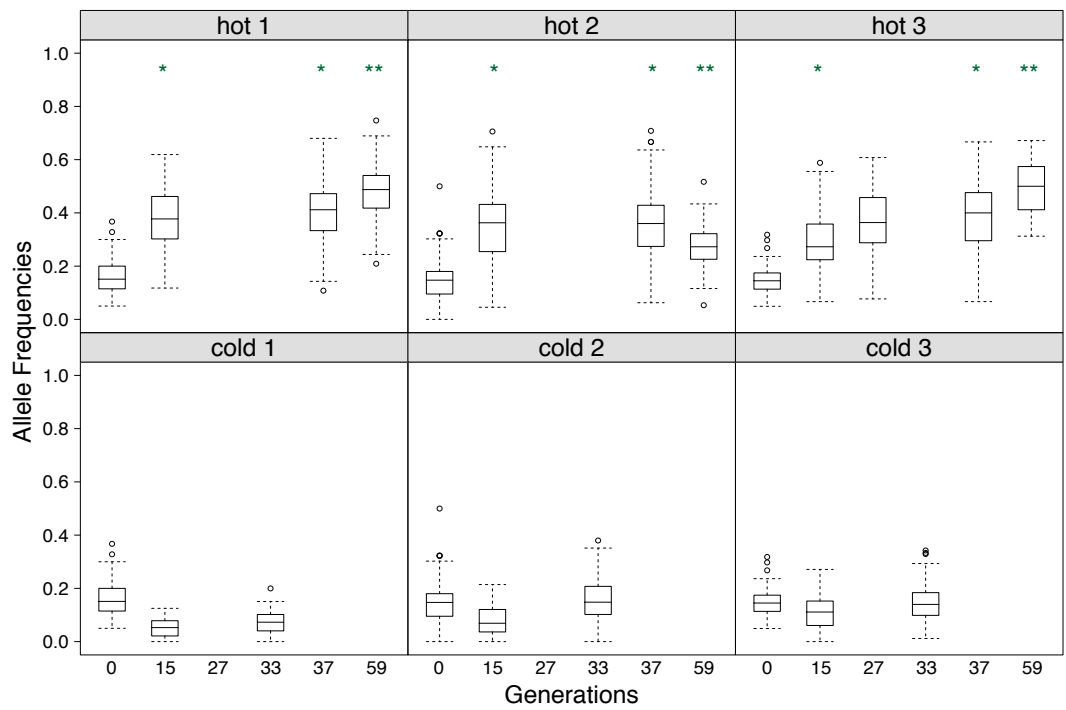

228

229

230

**E**

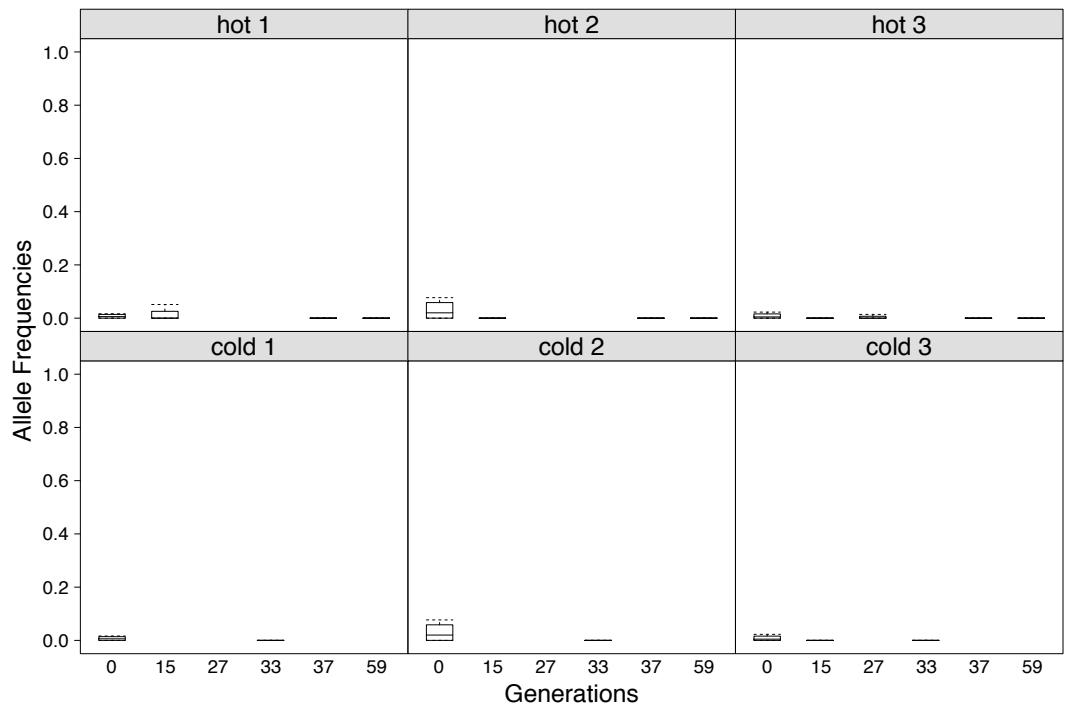

**F**

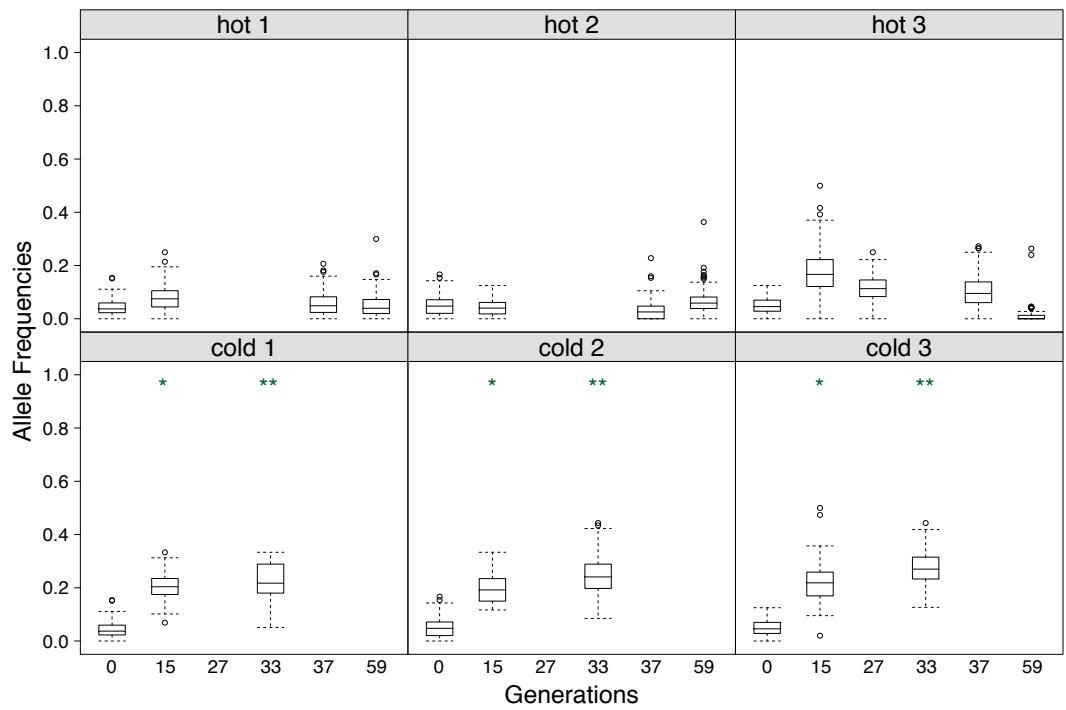

237    **G**

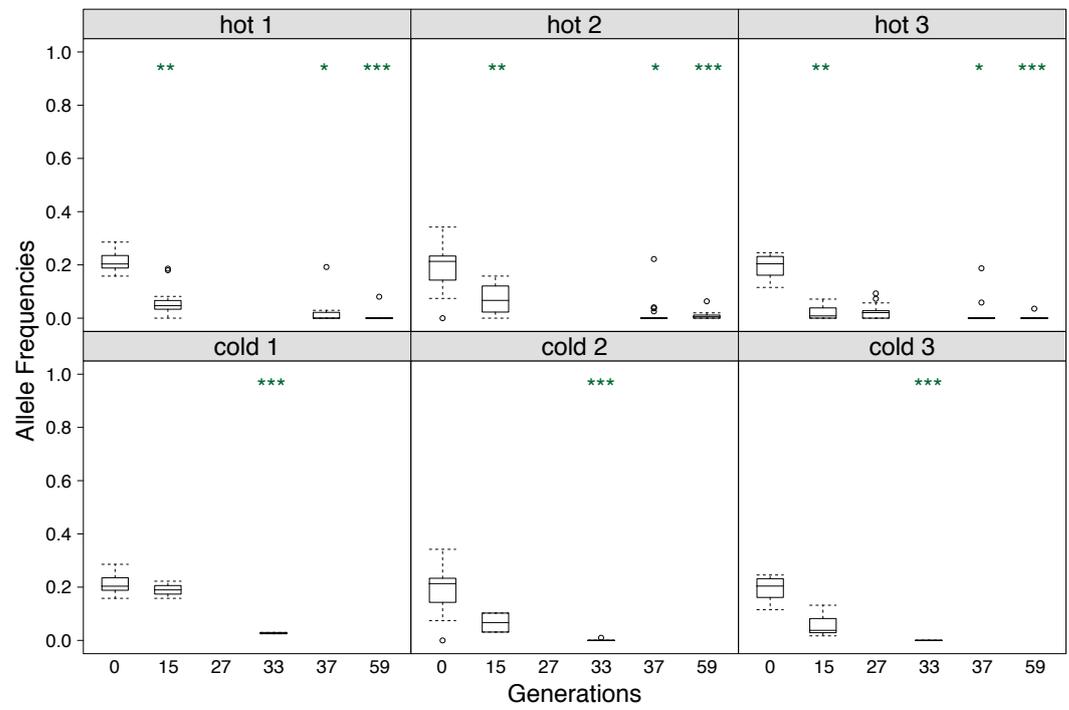

238

**Supporting Figure 4**

**A**

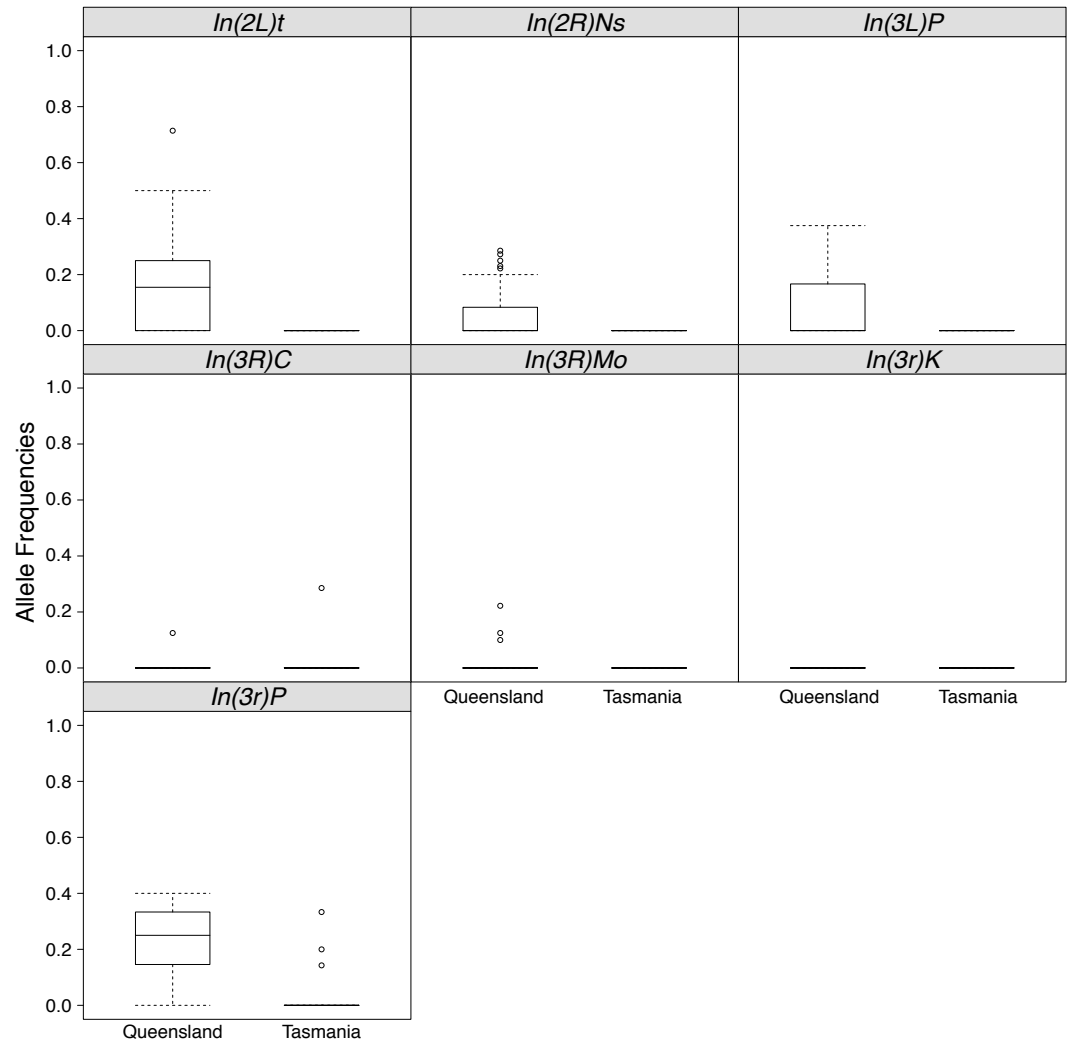

250  **B**

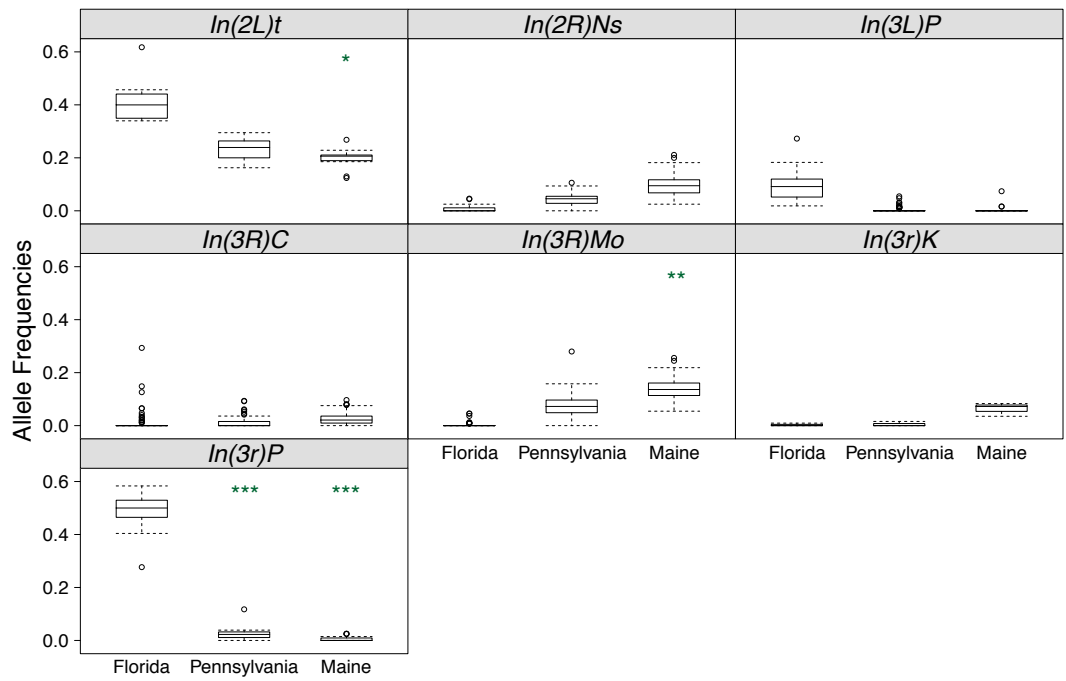

251

252

**Supporting Figure 5**

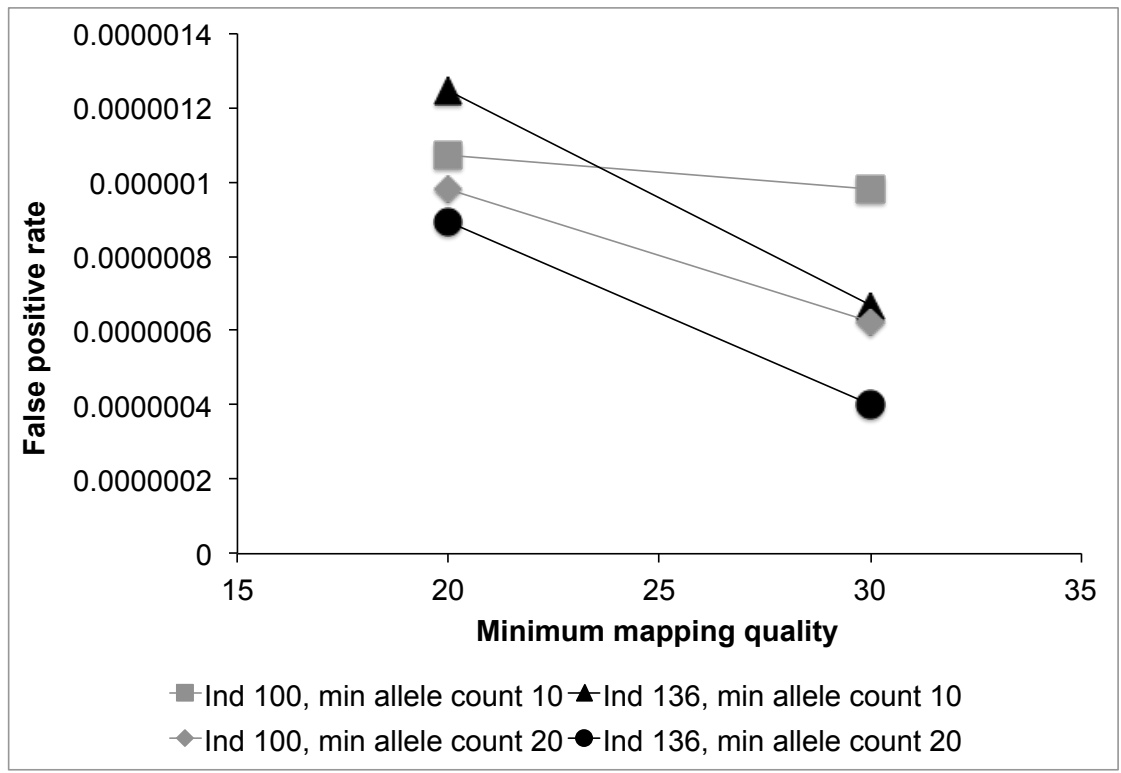

255 **Supporting Figure 6**

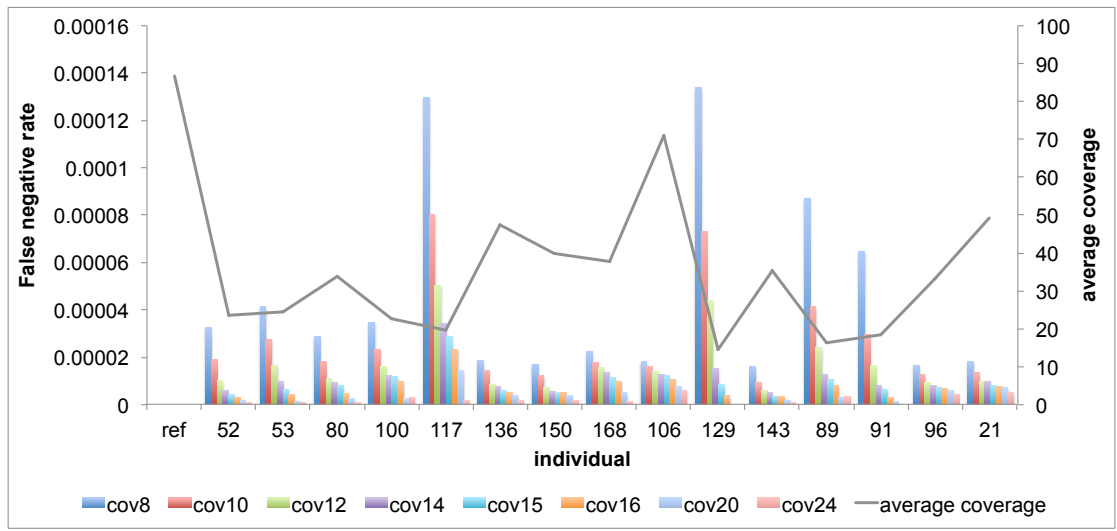

256

**Supporting Figure 7**

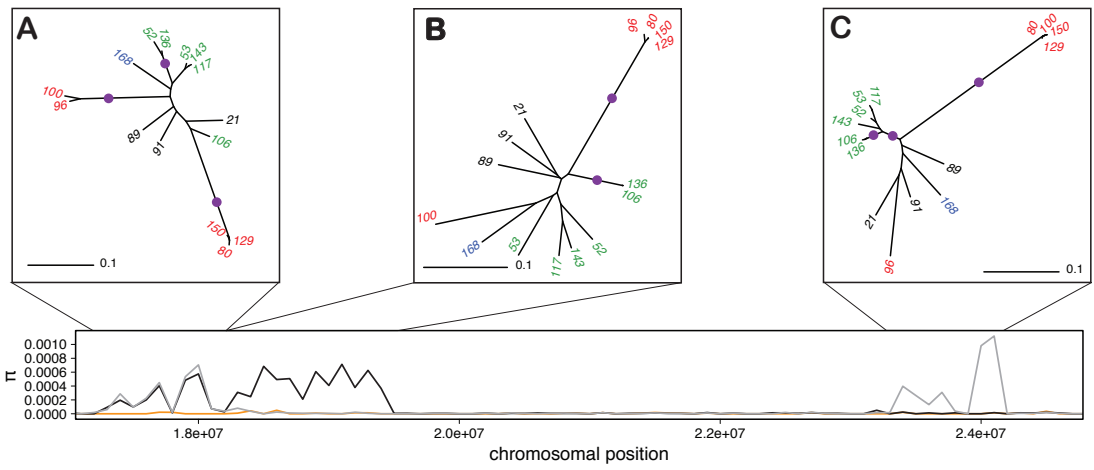

262 **Supporting Table 1. Karyotype and sex of sequenced individuals from the experimental evolution experiment.** Number of individual (ID),

263 selection regime ("hot", "cold"; replicates (R) 1-3), karyotype and sex of the 15 individuals sequenced from the experimental evolution

264 experiment. Also see Supporting Table 1 and Materials and Methods.

| ID | Regime | *In(2L)t* | *In(2R)Ns* | *In(3L)P* | *In(3R)C* | *In(3R)Mo* | *In(3R)P* | Sex |
|---|---|---|---|---|---|---|---|---|
| 21 | cold-R3 | 0 | 1 | 0 | 0 | 0 | 0 | f |
| 52 | cold-R2 | 0 | 0 | 0 | 1 | 0 | 0 | f |
| 53 | cold-R2 | 0 | 0 | 0 | 1 | 0 | 0 | f |
| 80 | cold-R2 | 1 | 0 | 0 | 0 | 1 | 0 | f |
| 89 | cold-R2 | 0 | 0 | 1 | 0 | 0 | 0 | f |
| 91 | cold-R1 | 0 | 0 | 1 | 0 | 0 | 0 | f |
| 96 | cold-R1 | 0 | 0 | 0 | 0 | 1 | 0 | f |
| 100 | cold-R1 | 1 | 0 | 0 | 0 | 1 | 0 | m |
| 106 | hot-R1 | 1 | 0 | 0 | 1 | 0 | 0 | f |
| 117 | hot-R1 | 0 | 0 | 1 | 1 | 0 | 0 | f |
| 129 | hot-R1 | 0 | 0 | 0 | 0 | 1 | 0 | f |
| 136 | hot-R1 | 1 | 0 | 0 | 1 | 0 | 0 | m |
| 143 | hot-R2 | 1 | 0 | 1 | 1 | 0 | 0 | f |
| 150 | hot-R2 | 0 | 0 | 0 | 0 | 1 | 0 | f |
| 168 | hot-R2 | 0 | 0 | 0 | 0 | 0 | 1 | f |

264  **Supporting Table 2. Individual karyotypes.** Data source, geographic origin, individual number (ID) and karyotype for all 167 individuals

265  used to identify fixed differences between chromosomal arrangements.

| Source | Origin | ID | *In(2L)t* | *In(2R)Ns* | *In(3L)P* | *In(3R)C* | *In(3R)K* | *In(3R)Mo* | *In(3R)P* |
|---|---|---|---|---|---|---|---|---|---|
| this study | Europe | 21 | 0 | 1 | 0 | 0 | 0 | 0 | 0 |
| this study | Europe | 52 | 0 | 0 | 0 | 1 | 0 | 0 | 0 |
| this study | Europe | 53 | 0 | 0 | 0 | 1 | 0 | 0 | 0 |
| this study | Europe | 80 | 1 | 0 | 0 | 0 | 0 | 1 | 0 |
| this study | Europe | 89 | 0 | 0 | 1 | 0 | 0 | 0 | 0 |
| this study | Europe | 91 | 0 | 0 | 1 | 0 | 0 | 0 | 0 |
| this study | Europe | 96 | 0 | 0 | 0 | 0 | 0 | 1 | 0 |
| this study | Europe | 100 | 1 | 0 | 0 | 0 | 0 | 1 | 0 |
| this study | Europe | 106 | 1 | 0 | 0 | 1 | 0 | 0 | 0 |
| this study | Europe | 117 | 0 | 0 | 1 | 1 | 0 | 0 | 0 |
| this study | Europe | 129 | 0 | 0 | 0 | 0 | 0 | 1 | 0 |
| this study | Europe | 136 | 1 | 0 | 0 | 1 | 0 | 0 | 0 |
| this study | Europe | 143 | 1 | 0 | 1 | 1 | 0 | 0 | 0 |
| this study | Europe | 150 | 0 | 0 | 0 | 0 | 0 | 1 | 0 |
| this study | Europe | 168 | 0 | 0 | 0 | 0 | 0 | 0 | 1 |
| DPGP2 | Africa | CK1 | 0 | 0 | 0 | 0 | 0 | 0 | 1 |
| DPGP2 | Africa | CK2 | 0 | 0 | 0 | 0 | 0 | 0 | 0 |
| DPGP2 | Africa | CO1 | 0 | 0 | 0 | 0 | 1 | 0 | 0 |
| DPGP2 | Africa | CO10N | 0 | 0 | 0 | 0 | 1 | 0 | 0 |
| DPGP2 | Africa | CO13N | 0 | 0 | 0 | 0 | 1 | 0 | 0 |
| DPGP2 | Africa | CO14 | 1 | 0 | 0 | 0 | 0 | 0 | 1 |
| DPGP2 | Africa | CO15N | 0 | 0 | 0 | 0 | 1 | 0 | 0 |
| DPGP2 | Africa | CO16 | 0 | 0 | 0 | 0 | 1 | 0 | 0 |
| DPGP2 | Africa | CO2 | 0 | 0 | 0 | 0 | 1 | 0 | 0 |
| DPGP2 | Africa | CO4N | 0 | 0 | 0 | 0 | 1 | 0 | 0 |
| DPGP2 | Africa | CO8N | 0 | 0 | 0 | 0 | 1 | 0 | 0 |
| DPGP2 | Africa | CO9N | 0 | 0 | 0 | 0 | 1 | 0 | 0 |
| DPGP2 | Africa | ED10N | 0 | 0 | 0 | 0 | 0 | 0 | 0 |
| DPGP2 | Africa | ED2 | 0 | 0 | 0 | 0 | 0 | 0 | 0 |
| DPGP2 | Africa | ED3 | 0 | 0 | 0 | 0 | 0 | 0 | 0 |
| DPGP2 | Africa | ED5N | 0 | 0 | 0 | 0 | 0 | 0 | 0 |
| DPGP2 | Africa | ED6N | 0 | 0 | 0 | 0 | 0 | 0 | 0 |
| DPGP2 | Africa | EZ2 | 1 | 0 | 0 | 0 | 0 | 0 | 0 |
| DPGP2 | Africa | EZ25 | 1 | 0 | 0 | 0 | 0 | 0 | 0 |
| DPGP2 | Africa | EZ5N | 0 | 0 | 0 | 0 | 0 | 0 | 0 |
| DPGP2 | Africa | EZ9N | 1 | 0 | 0 | 0 | 0 | 0 | 0 |
| DPGP2 | Europe | FR14 | 0 | 0 | 0 | 0 | 0 | 0 | 0 |
| DPGP2 | Europe | FR151 | 0 | 0 | 0 | 0 | 0 | 0 | 0 |
| DPGP2 | Europe | FR180 | 1 | 0 | 0 | 0 | 0 | 0 | 1 |
| DPGP2 | Europe | FR217 | 0 | 0 | 1 | 0 | 1 | 0 | 0 |
| DPGP2 | Europe | FR229 | 0 | 0 | 0 | 0 | 0 | 0 | 1 |
| DPGP2 | Europe | FR310 | 0 | 0 | 0 | 0 | 0 | 1 | 0 |
| DPGP2 | Europe | FR361 | 0 | 0 | 1 | 0 | 0 | 0 | 1 |
| DPGP2 | Europe | FR70 | 0 | 0 | 0 | 0 | 0 | 0 | 0 |
| DPGP2 | Africa | GA125 | 1 | 0 | 0 | 0 | 1 | 0 | 0 |
| DPGP2 | Africa | GA129 | 1 | 0 | 0 | 0 | 0 | 0 | 0 |
| DPGP2 | Africa | GA130 | 0 | 0 | 0 | 0 | 0 | 0 | 0 |
| DPGP2 | Africa | GA132 | 0 | 0 | 0 | 0 | 0 | 0 | 1 |
| DPGP2 | Africa | GA141 | 1 | 0 | 0 | 0 | 0 | 0 | 0 |
| DPGP2 | Africa | GA145  | 1 | 0 | 0 | 0 | 0 | 0 | 0 |
| DPGP2 | Africa | GA160  | 1 | 0 | 0 | 0 | 0 | 0 | 1 |
| DPGP2 | Africa | GA185  | 0 | 0 | 0 | 0 | 0 | 0 | 1 |
| DPGP2 | Africa | GA191  | 0 | 0 | 0 | 0 | 0 | 0 | 0 |
| DPGP2 | Africa | GU10   | 0 | 0 | 0 | 0 | 1 | 0 | 0 |
| DPGP2 | Africa | GU2    | 0 | 0 | 0 | 0 | 0 | 0 | 0 |
| DPGP2 | Africa | GU6    | 0 | 0 | 0 | 0 | 1 | 0 | 0 |
| DPGP2 | Africa | GU7    | 1 | 0 | 0 | 0 | 0 | 0 | 1 |
| DPGP2 | Africa | GU9    | 0 | 0 | 0 | 0 | 0 | 0 | 1 |
| DPGP2 | Africa | KN133N | 1 | 0 | 0 | 0 | 0 | 0 | 0 |
| DPGP2 | Africa | KN20N  | 0 | 0 | 0 | 0 | 1 | 0 | 0 |
| DPGP2 | Africa | KN34   | 0 | 0 | 0 | 0 | 0 | 0 | 0 |
| DPGP2 | Africa | KN35   | 1 | 0 | 0 | 0 | 0 | 0 | 0 |
| DPGP2 | Africa | KN6    | 0 | 0 | 0 | 0 | 0 | 0 | 0 |
| DPGP2 | Africa | KR39   | 1 | 0 | 0 | 0 | 0 | 0 | 0 |
| DPGP2 | Africa | KR42   | 1 | 1 | 0 | 0 | 0 | 0 | 0 |
| DPGP2 | Africa | KR4N   | 1 | 0 | 0 | 0 | 0 | 0 | 0 |
| DPGP2 | Africa | KR7    | 1 | 0 | 0 | 0 | 0 | 0 | 0 |
| DPGP2 | Africa | KT1    | 0 | 0 | 0 | 0 | 0 | 0 | 0 |
| DPGP2 | Africa | KT6    | 0 | 1 | 0 | 0 | 0 | 0 | 0 |
| DPGP2 | Africa | NG10N  | 1 | 0 | 1 | 0 | 0 | 0 | 1 |
| DPGP2 | Africa | NG1N   | 1 | 1 | 0 | 0 | 1 | 0 | 0 |
| DPGP2 | Africa | NG3N   | 1 | 0 | 1 | 0 | 0 | 0 | 0 |
| DPGP2 | Africa | NG6N   | 0 | 0 | 0 | 0 | 0 | 0 | 0 |
| DPGP2 | Africa | NG7    | 0 | 0 | 0 | 0 | 0 | 0 | 0 |
| DPGP2 | Africa | NG9    | 1 | 1 | 1 | 0 | 0 | 0 | 0 |
| DPGP2 | Africa | RC1    | 0 | 0 | 0 | 0 | 0 | 0 | 0 |
| DPGP2 | Africa | RC5   | 0 | 0 | 0 | 0 | 0 | 0 | 0 |
| DPGP2 | Africa | RG10  | 0 | 0 | 0 | 0 | 0 | 0 | 0 |
| DPGP2 | Africa | RG11N | 0 | 0 | 0 | 0 | 0 | 0 | 0 |
| DPGP2 | Africa | RG13N | 0 | 0 | 0 | 0 | 0 | 0 | 0 |
| DPGP2 | Africa | RG15  | 0 | 0 | 0 | 0 | 0 | 0 | 0 |
| DPGP2 | Africa | RG18N | 0 | 0 | 0 | 0 | 0 | 0 | 1 |
| DPGP2 | Africa | RG19  | 0 | 0 | 0 | 0 | 0 | 0 | 0 |
| DPGP2 | Africa | RG2   | 0 | 0 | 0 | 0 | 0 | 0 | 0 |
| DPGP2 | Africa | RG21N | 0 | 0 | 0 | 0 | 0 | 0 | 0 |
| DPGP2 | Africa | RG22  | 0 | 0 | 0 | 0 | 0 | 0 | 0 |
| DPGP2 | Africa | RG24  | 0 | 0 | 0 | 0 | 0 | 0 | 0 |
| DPGP2 | Africa | RG25  | 0 | 0 | 0 | 0 | 0 | 0 | 1 |
| DPGP2 | Africa | RG28  | 0 | 0 | 0 | 0 | 0 | 0 | 0 |
| DPGP2 | Africa | RG3   | 1 | 1 | 0 | 0 | 0 | 0 | 0 |
| DPGP2 | Africa | RG32N | 0 | 0 | 0 | 0 | 0 | 0 | 0 |
| DPGP2 | Africa | RG33  | 0 | 0 | 0 | 0 | 0 | 0 | 0 |
| DPGP2 | Africa | RG34  | 0 | 0 | 0 | 0 | 0 | 0 | 0 |
| DPGP2 | Africa | RG35  | 0 | 0 | 0 | 0 | 0 | 0 | 0 |
| DPGP2 | Africa | RG36  | 1 | 0 | 0 | 0 | 0 | 0 | 0 |
| DPGP2 | Africa | RG37N | 1 | 0 | 0 | 0 | 0 | 0 | 0 |
| DPGP2 | Africa | RG38N | 0 | 0 | 0 | 0 | 0 | 0 | 0 |
| DPGP2 | Africa | RG39  | 0 | 0 | 0 | 0 | 0 | 0 | 0 |
| DPGP2 | Africa | RG4N  | 0 | 0 | 0 | 0 | 0 | 0 | 0 |
| DPGP2 | Africa | RG5   | 0 | 0 | 0 | 0 | 0 | 0 | 1 |
| DPGP2 | Africa | RG6N  | 0 | 0 | 0 | 0 | 0 | 0 | 0 |
| DPGP2 | Africa | RG7   | 0 | 0 | 0 | 0 | 0 | 0 | 0 |
| DPGP2 | Africa | RG8   | 0 | 0 | 0 | 0 | 0 | 0 | 0 |
| DPGP2 | Africa | RG9    | 0 | 0 | 0 | 0 | 0 | 0 | 1 |
| DPGP2 | Africa | SP173  | 0 | 0 | 0 | 0 | 0 | 0 | 0 |
| DPGP2 | Africa | SP188  | 0 | 0 | 0 | 0 | 0 | 0 | 0 |
| DPGP2 | Africa | SP221  | 1 | 1 | 0 | 0 | 0 | 0 | 0 |
| DPGP2 | Africa | SP235  | 0 | 0 | 0 | 0 | 0 | 0 | 0 |
| DPGP2 | Africa | SP241  | 0 | 0 | 0 | 0 | 0 | 0 | 0 |
| DPGP2 | Africa | SP254  | 0 | 0 | 0 | 0 | 0 | 0 | 0 |
| DPGP2 | Africa | SP80   | 0 | 0 | 0 | 0 | 0 | 0 | 0 |
| DPGP2 | Africa | TZ10   | 1 | 0 | 0 | 0 | 0 | 0 | 0 |
| DPGP2 | Africa | TZ14   | 1 | 0 | 0 | 0 | 1 | 0 | 0 |
| DPGP2 | Africa | TZ8    | 1 | 0 | 0 | 0 | 0 | 0 | 0 |
| DPGP2 | Africa | UG19   | 0 | 0 | 0 | 0 | 0 | 0 | 0 |
| DPGP2 | Africa | UG28N  | 0 | 0 | 0 | 0 | 0 | 0 | 0 |
| DPGP2 | Africa | UG5N   | 0 | 0 | 0 | 0 | 0 | 0 | 0 |
| DPGP2 | Africa | UG7    | 0 | 0 | 0 | 0 | 0 | 0 | 0 |
| DPGP2 | Africa | UM118  | 1 | 0 | 0 | 0 | 0 | 0 | 0 |
| DPGP2 | Africa | UM37   | 0 | 0 | 0 | 0 | 1 | 0 | 0 |
| DPGP2 | Africa | UM526  | 1 | 0 | 0 | 0 | 0 | 0 | 0 |
| DPGP2 | Africa | ZI261  | 0 | 0 | 0 | 0 | 0 | 0 | 0 |
| DPGP2 | Africa | ZI268  | 0 | 0 | 0 | 0 | 0 | 0 | 0 |
| DPGP2 | Africa | ZI468  | 0 | 0 | 0 | 0 | 0 | 0 | 0 |
| DPGP2 | Africa | ZI91   | 0 | 0 | 0 | 0 | 0 | 0 | 0 |
| DPGP2 | Africa | ZL130  | 0 | 0 | 0 | 0 | 0 | 0 | 0 |
| DPGP2 | Africa | ZO65   | 1 | 1 | 0 | 0 | 0 | 0 | 0 |
| DPGP2 | Africa | ZS11   | 1 | 0 | 0 | 0 | 0 | 0 | 0 |
| DPGP2 | Africa | ZS37   | 0 | 0 | 0 | 0 | 1 | 0 | 0 |
| DPGP2 | Africa | ZS5    | 0 | 1 | 0 | 0 | 0 | 0 | 0 |
| DPGP | North America | RAL-301 | 1 | 0 | 0 | 0 | 0 | 0 | 0 |
| DPGP | North America | RAL-303 | 0 | 0 | 0 | 0 | 0 | 0 | 0 |
| DPGP | North America | RAL-304 | 0 | 1 | 0 | 0 | 0 | 0 | 0 |
| DPGP | North America | RAL-306 | 0 | 0 | 0 | 0 | 0 | 0 | 0 |
| DPGP | North America | RAL-307 | 0 | 0 | 0 | 0 | 0 | 0 | 0 |
| DPGP | North America | RAL-313 | 1 | 0 | 0 | 0 | 0 | 0 | 0 |
| DPGP | North America | RAL-315 | 0 | 0 | 0 | 0 | 0 | 0 | 0 |
| DPGP | North America | RAL-324 | 0 | 0 | 0 | 0 | 0 | 1 | 0 |
| DPGP | North America | RAL-335 | 0 | 0 | 0 | 0 | 0 | 0 | 0 |
| DPGP | North America | RAL-357 | 0 | 0 | 0 | 0 | 0 | 0 | 0 |
| DPGP | North America | RAL-358 | 1 | 0 | 0 | 0 | 0 | 1 | 0 |
| DPGP | North America | RAL-360 | 0 | 0 | 0 | 0 | 0 | 0 | 0 |
| DPGP | North America | RAL-362 | 0 | 0 | 0 | 0 | 0 | 0 | 0 |
| DPGP | North America | RAL-365 | 0 | 0 | 0 | 0 | 0 | 0 | 0 |
| DPGP | North America | RAL-375 | 0 | 0 | 0 | 0 | 0 | 0 | 0 |
| DPGP | North America | RAL-379 | 0 | 0 | 0 | 0 | 0 | 0 | 0 |
| DPGP | North America | RAL-380 | 0 | 0 | 0 | 0 | 0 | 0 | 0 |
| DPGP | North America | RAL-391 | 0 | 0 | 0 | 0 | 0 | 0 | 0 |
| DPGP | North America | RAL-399 | 0 | 0 | 0 | 0 | 0 | 0 | 0 |
| DPGP | North America | RAL-427 | 0 | 0 | 0 | 0 | 0 | 0 | 0 |
| DPGP | North America | RAL-437 | 0 | 0 | 0 | 0 | 0 | 1 | 0 |
| DPGP | North America | RAL-486 | 0 | 0 | 0 | 0 | 0 | 0 | 0 |
| DPGP | North America | RAL-514 | 0 | 0 | 0 | 0 | 0 | 0 | 0 |
| DPGP | North America | RAL-517 | 0 | 0 | 0 | 0 | 0 | 0 | 0 |
| DPGP | North America | RAL-555 | 0 | 0 | 0 | 0 | 0 | 1 | 0 |
| DPGP | North America | RAL-639 | 0 | 0 | 0 | 0 | 0 | 0 | 0 |
| DPGP | North America | RAL-705 | 0 | 0 | 0 | 0 | 0 | 0 | 0 |
| DPGP | North America | RAL-707 | 0 | 0 | 0 | 0 | 0 | 1 | 0 |
| DPGP | North America | RAL-714 | 0 | 0 | 0 | 0 | 0 | 1 | 0 |
| DPGP | North America | RAL-730 | 0 | 0 | 0 | 0 | 0 | 0 | 0 |
| DPGP | North America | RAL-732 | 0 | 0 | 0 | 0 | 1 | 0 | 0 |
| DPGP | North America | RAL-765 | 0 | 0 | 0 | 0 | 0 | 0 | 0 |
| DPGP | North America | RAL-774 | 0 | 0 | 0 | 0 | 0 | 0 | 0 |
| DPGP | North America | RAL-786 | 0 | 0 | 0 | 0 | 0 | 0 | 1 |
| DPGP | North America | RAL-799 | 0 | 0 | 0 | 0 | 0 | 0 | 0 |
| DPGP | North America | RAL-820 | 0 | 0 | 0 | 0 | 0 | 1 | 0 |
| DPGP | North America | RAL-852 | 0 | 1 | 0 | 0 | 0 | 0 | 0 |

**Supporting Table 3. Inversion-specific marker alleles.** Chromosomal position and inversion-specific allele for the fixed differences between the corresponding inversion and all other chromosomal arrangements, based on 167 chromosomes.

| Inversion | Chromosome | Position | Allele |
|---|---|---|---|
| *In(2L)t* | 2L | 2166548 | A |
| *In(2L)t* | 2L | 2166622 | G |
| *In(2L)t* | 2L | 2166626 | A |
| *In(2L)t* | 2L | 2204678 | A |
| *In(2L)t* | 2L | 2209048 | C |
| *In(2L)t* | 2L | 2214322 | T |
| *In(2L)t* | 2L | 2225369 | T |
| *In(2L)t* | 2L | 2226971 | G |
| *In(2L)t* | 2L | 2233906 | A |
| *In(2L)t* | 2L | 2234101 | A |
| *In(2L)t* | 2L | 2246686 | T |
| *In(2L)t* | 2L | 2255218 | A |
| *In(2L)t* | 2L | 13139098 | C |
| *In(2L)t* | 2L | 13155257 | T |
| *In(2L)t* | 2L | 13172139 | T |
| *In(2L)t* | 2L | 13186585 | A |
| *In(2R)Ns* | 2R | 11279637 | A |
| *In(2R)Ns* | 2R | 11291326 | A |
| *In(2R)Ns* | 2R | 11291656 | A |
| *In(2R)Ns* | 2R | 11294553 | A |
| *In(2R)Ns* | 2R | 11295105 | A |
| *In(2R)Ns* | 2R | 11295408 | A |
| *In(2R)Ns* | 2R | 11297771 | T |
| *In(2R)Ns* | 2R | 11298425 | C |
| *In(2R)Ns* | 2R | 11363601 | T |
| *In(2R)Ns* | 2R | 11416627 | T |
| *In(2R)Ns* | 2R | 11416743 | G |
| *In(2R)Ns* | 2R | 11428502 | G |
| *In(2R)Ns* | 2R | 11452011 | C |
| *In(2R)Ns* | 2R | 11453509 | T |
| *In(2R)Ns* | 2R | 11459978 | G |
| *In(2R)Ns* | 2R | 11467228 | T |
| *In(2R)Ns* | 2R | 11470424 | T |
| *In(2R)Ns* | 2R | 11471637 | T |
| *In(2R)Ns* | 2R | 11620344 | A |
| *In(2R)Ns* | 2R | 11685989 | T |
| *In(2R)Ns* | 2R | 11817613 | A |
| *In(2R)Ns* | 2R | 11818383 | T |
| *In(2R)Ns* | 2R | 11826149 | T |
| *In(2R)Ns* | 2R | 12007749 | A |
| *In(2R)Ns* | 2R | 12154859 | A |
| *In(2R)Ns* | 2R | 12250521 | T |
| *In(2R)Ns* | 2R | 12394846 | G |
| *In(2R)Ns* | 2R | 13942780 | A |
| *In(2R)Ns* | 2R | 13944397 | C |
| *In(2R)Ns* | 2R | 14352759 | A |
| *In(2R)Ns* | 2R | 14362949 | T |
| *In(2R)Ns* | 2R | 14582447 | T |
| *In(2R)Ns* | 2R | 14633978 | T |
| *In(2R)Ns* | 2R | 14641278 | A |
| *In(2R)Ns* | 2R | 14672926 | A |
| *In(2R)Ns* | 2R | 14674348 | T |
| *In(2R)Ns* | 2R | 14735385 | G |
| *In(2R)Ns* | 2R | 14995376 | T |
| *In(2R)Ns* | 2R | 15117841 | T |
| *In(2R)Ns* | 2R | 15122558 | G |
| *In(2R)Ns* | 2R | 15124138 | T |
| *In(2R)Ns* | 2R | 15154801 | A |
| *In(2R)Ns* | 2R | 15160191 | A |
| *In(2R)Ns* | 2R | 15289938 | G |
| *In(2R)Ns* | 2R | 15303213 | T |
| *In(2R)Ns* | 2R | 15303225 | A |
| *In(2R)Ns* | 2R | 15335793 | T |
| *In(2R)Ns* | 2R | 15339141 | T |
| *In(2R)Ns* | 2R | 15339337 | T |
| *In(2R)Ns* | 2R | 15344384 | A |
| *In(2R)Ns* | 2R | 15345300 | T |
| *In(2R)Ns* | 2R | 15348825 | A |
| *In(2R)Ns* | 2R | 15364662 | C |
| *In(2R)Ns* | 2R | 15364670 | C |
| *In(2R)Ns* | 2R | 15366984 | A |
| *In(2R)Ns* | 2R | 15367369 | A |
| *In(2R)Ns* | 2R | 15370164 | A |
| *In(2R)Ns* | 2R | 16023748 | T |
| *In(2R)Ns* | 2R | 16071561 | T |
| *In(2R)Ns* | 2R | 16073117 | T |
| *In(2R)Ns* | 2R | 16100012 | T |
| *In(2R)Ns* | 2R | 16116600 | A |
| *In(2R)Ns* | 2R | 16117724 | T |
| *In(2R)Ns* | 2R | 16152311 | C |
| *In(2R)Ns* | 2R | 16152687 | G |
| *In(2R)Ns* | 2R | 16160042 | T |
| *In(2R)Ns* | 2R | 16163328 | T |
| *In(3L)P* | 3L | 2759715 | C |
| *In(3L)P* | 3L | 2760784 | T |
| *In(3L)P* | *3L* | 3054925 | C |
| *In(3L)P* | *3L* | 3133022 | G |
| *In(3L)P* | *3L* | 3135682 | C |
| *In(3L)P* | *3L* | 3142231 | A |
| *In(3L)P* | *3L* | 3145702 | T |
| *In(3L)P* | *3L* | 3148304 | A |
| *In(3L)P* | *3L* | 3152282 | C |
| *In(3L)P* | *3L* | 3156337 | A |
| *In(3L)P* | *3L* | 3165913 | A |
| *In(3L)P* | *3L* | 3172232 | A |
| *In(3L)P* | *3L* | 3172572 | A |
| *In(3L)P* | *3L* | 3190585 | G |
| *In(3L)P* | *3L* | 3191474 | G |
| *In(3L)P* | *3L* | 3192621 | A |
| *In(3L)P* | *3L* | 3194000 | T |
| *In(3L)P* | *3L* | 3195095 | A |
| *In(3L)P* | *3L* | 3198656 | T |
| *In(3L)P* | *3L* | 3202276 | G |
| *In(3L)P* | *3L* | 3203140 | T |
| *In(3L)P* | *3L* | 3203449 | G |
| *In(3L)P* | *3L* | 3205464 | C |
| *In(3L)P* | *3L* | 3244232 | A |
| *In(3L)P* | *3L* | 3250267 | C |
| *In(3L)P* | *3L* | 3251643 | G |
| *In(3L)P* | *3L* | 3258888 | A |
| *In(3L)P* | *3L* | 3260348 | A |
| *In(3L)P* | *3L* | 3274254 | C |
| *In(3L)P* | *3L* | 3284533 | A |
| *In(3L)P* | *3L* | 3388479 | G |
| *In(3L)P* | *3L* | 3389696 | A |
| *In(3L)P* | *3L* | 3390222 | G |
| *In(3L)P* | *3L* | 3397051 | A |
| *In(3L)P* | *3L* | 3430131 | G |
| *In(3L)P* | *3L* | 3764444 | T |
| *In(3L)P* | *3L* | 5399565 | T |
| *In(3L)P* | *3L* | 15633845 | G |
| *In(3L)P* | *3L* | 15970961 | G |
| *In(3L)P* | *3L* | 16165187 | A |
| *In(3L)P* | *3L* | 16165189 | T |
| *In(3L)P* | *3L* | 16165230 | C |
| *In(3L)P* | *3L* | 16170296 | C |
| *In(3L)P* | *3L* | 16193541 | A |
| *In(3L)P* | *3L* | 16201506 | G |
| *In(3L)P* | *3L* | 16217175 | A |
| *In(3L)P* | *3L* | 16222536 | G |
| *In(3L)P* | *3L* | 16223154 | C |
| *In(3L)P* | *3L* | 16261646 | T |
| *In(3L)P* | *3L* | 16261672 | A |
| *In(3L)P* | *3L* | 16261695 | T |
| *In(3L)P* | *3L* | 16261726 | T |
| *In(3L)P* | *3L* | 16263247 | C |
| *In(3L)P* | *3L* | 16263588 | T |
| *In(3L)P* | *3L* | 16268717 | T |
| *In(3L)P* | *3L* | 16273480 | G |
| *In(3L)P* | *3L* | 16280796 | G |
| *In(3L)P* | *3L* | 16280798 | A |
| *In(3L)P* | *3L* | 16289482 | C |
| *In(3L)P* | *3L* | 16290594 | G |
| *In(3L)P* | *3L* | 16290972 | T |
| *In(3L)P* | *3L* | 16291332 | G |
| *In(3L)P* | *3L* | 16297916 | A |
| *In(3L)P* | *3L* | 16298085 | A |
| *In(3L)P* | *3L* | 16301520 | A |
| *In(3L)P* | *3L* | 16308563 | C |
| *In(3L)P* | *3L* | 16311425 | C |
| *In(3L)P* | *3L* | 16326362 | T |
| *In(3L)P* | *3L* | 16333526 | A |
| *In(3L)P* | *3L* | 16377449 | A |
| *In(3L)P* | *3L* | 16378572 | T |
| *In(3L)P* | *3L* | 16393822 | C |
| *In(3L)P* | *3L* | 16400709 | G |
| *In(3R)C* | *3R* | 13114726 | T |
| *In(3R)C* | *3R* | 16099151 | G |
| *In(3R)C* | *3R* | 16104479 | A |
| *In(3R)C* | *3R* | 16110028 | T |
| *In(3R)C* | *3R* | 16114832 | G |
| *In(3R)C* | *3R* | 16145902 | G |
| *In(3R)C* | *3R* | 16145903 | T |
| *In(3R)C* | *3R* | 16191928 | T |
| *In(3R)C* | *3R* | 16864615 | C |
| *In(3R)C* | *3R* | 16893226 | C |
| *In(3R)C* | *3R* | 16918188 | T |
| *In(3R)C* | *3R* | 19748559 | A |
| *In(3R)C* | *3R* | 19755935 | T |
| *In(3R)C* | *3R* | 20442534 | G |
| *In(3R)C* | *3R* | 20498606 | G |
| *In(3R)C* | *3R* | 20558459 | G |
| *In(3R)C* | *3R* | 20924283 | T |
| *In(3R)C* | *3R* | 20943910 | T |
| *In(3R)C* | *3R* | 23033890 | A |
| *In(3R)C* | *3R* | 24007045 | T |
| *In(3R)C* | *3R* | 24007371 | G |
| *In(3R)C* | *3R* | 24009461 | T |
| *In(3R)C* | *3R* | 24014066 | T |
| *In(3R)C* | *3R* | 24029634 | T |
| *In(3R)C* | *3R* | 24041884 | A |
| *In(3R)C* | *3R* | 24041990 | T |
| *In(3R)C* | *3R* | 24043681 | C |
| *In(3R)C* | *3R* | 24044393 | A |
| *In(3R)C* | *3R* | 24078020 | G |
| *In(3R)C* | *3R* | 24085873 | T |
| *In(3R)C* | *3R* | 24096291 | T |
| *In(3R)C* | *3R* | 24138943 | G |
| *In(3R)C* | *3R* | 24142235 | C |
| *In(3R)C* | *3R* | 24150589 | T |
| *In(3R)C* | *3R* | 24163991 | T |
| *In(3R)C* | *3R* | 24171563 | A |
| *In(3R)C* | *3R* | 24172382 | A |
| *In(3R)C* | *3R* | 24195591 | G |
| *In(3R)C* | *3R* | 24201208 | T |
| *In(3R)C* | *3R* | 24242753 | A |
| *In(3R)C* | *3R* | 24243280 | C |
| *In(3R)C* | *3R* | 24279617 | G |
| *In(3R)C* | *3R* | 24282605 | T |
| *In(3R)C* | *3R* | 24298461 | A |
| *In(3R)C* | *3R* | 24342811 | T |
| *In(3R)C* | *3R* | 24374212 | G |
| *In(3R)C* | *3R* | 24409151 | A |
| *In(3R)C* | *3R* | 24422474 | C |
| *In(3R)C* | *3R* | 24467871 | T |
| *In(3R)C* | *3R* | 24487712 | G |
| *In(3R)C* | *3R* | 24493367 | G |
| *In(3R)C* | *3R* | 24506558 | G |
| *In(3R)C* | *3R* | 24512937 | T |
| *In(3R)C* | *3R* | 24522397 | G |
| *In(3R)C* | *3R* | 24551095 | A |
| *In(3R)C* | *3R* | 24690673 | T |
| *In(3R)C* | *3R* | 24693933 | A |
| *In(3R)C* | *3R* | 24694365 | A |
| *In(3R)C* | *3R* | 24719313 | A |
| *In(3R)C* | *3R* | 25096252 | A |
| *In(3R)C* | *3R* | 25106453 | C |
| *In(3R)C* | *3R* | 25136719 | A |
| *In(3R)C* | *3R* | 25175337 | A |
| *In(3R)C* | *3R* | 25176234 | G |
| *In(3R)C* | *3R* | 25179516 | G |
| *In(3R)C* | *3R* | 25193278 | A |
| *In(3R)C* | *3R* | 25216865 | A |

| *In(3R)C* | 3R | 25222529 | G |
| *In(3R)C* | 3R | 25242597 | G |
| *In(3R)C* | 3R | 25248195 | T |
| *In(3R)C* | 3R | 25269879 | A |
| *In(3R)C* | 3R | 25315158 | A |
| *In(3R)C* | 3R | 25329587 | C |
| *In(3R)C* | 3R | 25474612 | T |
| *In(3R)C* | 3R | 25489586 | C |
| *In(3R)C* | 3R | 25505585 | C |
| *In(3R)C* | 3R | 25538313 | A |
| *In(3R)C* | 3R | 25560925 | A |
| *In(3R)C* | 3R | 25567683 | C |
| *In(3R)C* | 3R | 25583469 | A |
| *In(3R)C* | 3R | 25596484 | T |
| *In(3R)C* | 3R | 25598648 | C |
| *In(3R)C* | 3R | 25599170 | T |
| *In(3R)C* | 3R | 25604540 | T |
| *In(3R)C* | 3R | 25604725 | C |
| *In(3R)C* | 3R | 25605392 | G |
| *In(3R)C* | 3R | 25605428 | T |
| *In(3R)C* | 3R | 25632833 | A |
| *In(3R)C* | 3R | 25647947 | C |
| *In(3R)C* | 3R | 25680387 | G |
| *In(3R)C* | 3R | 25686401 | C |
| *In(3R)C* | 3R | 25686744 | A |
| *In(3R)C* | 3R | 25689415 | G |
| *In(3R)C* | 3R | 25689478 | T |
| *In(3R)C* | 3R | 25692175 | T |
| *In(3R)C* | 3R | 25776627 | C |
| *In(3R)C* | 3R | 25789208 | A |
| *In(3R)C* | 3R | 25789641 | C |
| *In(3R)C* | 3R | 25798811 | A |
| *In(3R)C* | 3R | 25810959 | T |
| *In(3R)C* | 3R | 25822138 | T |
| *In(3R)C* | 3R | 25830799 | T |
| *In(3R)C* | 3R | 25836339 | T |
| *In(3R)C* | 3R | 25865969 | A |
| *In(3R)C* | 3R | 25881149 | A |
| *In(3R)C* | 3R | 25884722 | G |
| *In(3R)C* | 3R | 25885398 | A |
| *In(3R)C* | 3R | 25885568 | A |
| *In(3R)C* | 3R | 25892882 | T |
| *In(3R)C* | 3R | 25893312 | T |
| *In(3R)C* | 3R | 25901563 | T |
| *In(3R)C* | 3R | 25904049 | C |
| *In(3R)C* | 3R | 25904085 | G |
| *In(3R)C* | *3R* | 26052763 | T |
| *In(3R)C* | *3R* | 26450277 | C |
| *In(3R)C* | *3R* | 26502830 | A |
| *In(3R)C* | *3R* | 26541828 | C |
| *In(3R)C* | *3R* | 26553123 | T |
| *In(3R)C* | *3R* | 26833261 | T |
| *In(3R)C* | *3R* | 27033799 | A |
| *In(3R)C* | *3R* | 27050399 | C |
| *In(3R)C* | *3R* | 27050401 | G |
| *In(3R)C* | *3R* | 27183127 | A |
| *In(3R)C* | *3R* | 27187114 | G |
| *In(3R)C* | *3R* | 27189512 | G |
| *In(3R)C* | *3R* | 27213181 | T |
| *In(3R)C* | *3R* | 27230179 | G |
| *In(3R)C* | *3R* | 27255032 | G |
| *In(3R)C* | *3R* | 27348805 | A |
| *In(3R)C* | *3R* | 27350380 | T |
| *In(3R)C* | *3R* | 27355100 | A |
| *In(3R)C* | *3R* | 27355101 | T |
| *In(3R)C* | *3R* | 27367655 | T |
| *In(3R)C* | *3R* | 27376219 | A |
| *In(3R)C* | *3R* | 27450892 | T |
| *In(3R)C* | *3R* | 27536048 | G |
| *In(3R)C* | *3R* | 27560508 | G |
| *In(3R)C* | *3R* | 27560856 | A |
| *In(3R)C* | *3R* | 27561118 | A |
| *In(3R)C* | *3R* | 27813043 | T |
| *In(3R)C* | *3R* | 27815314 | C |
| *In(3R)C* | *3R* | 27819657 | C |
| *In(3R)C* | *3R* | 27873302 | A |
| *In(3R)C* | *3R* | 27885889 | A |
| *In(3R)K* | *3R* | 7569591 | G |
| *In(3R)K* | *3R* | 7587158 | A |
| *In(3R)K* | *3R* | 7763547 | T |
| *In(3R)K* | *3R* | 21961212 | C |
| *In(3R)Mo* | *3R* | 15955370 | C |
| *In(3R)Mo* | *3R* | 15956205 | G |
| *In(3R)Mo* | *3R* | 16012652 | A |
| *In(3R)Mo* | *3R* | 16054389 | T |
| *In(3R)Mo* | *3R* | 16088352 | T |
| *In(3R)Mo* | *3R* | 16101901 | A |
| *In(3R)Mo* | *3R* | 16309968 | A |
| *In(3R)Mo* | *3R* | 16310458 | T |
| *In(3R)Mo* | *3R* | 16321720 | G |
| *In(3R)Mo* | *3R* | 16324886 | T |
| *In(3R)Mo* | *3R* | 16327977 | A |
| *In(3R)Mo* | *3R* | 16329725 | C |
| *In(3R)Mo* | *3R* | 16354768 | T |
| *In(3R)Mo* | *3R* | 16358463 | A |
| *In(3R)Mo* | *3R* | 16477118 | A |
| *In(3R)Mo* | *3R* | 16505890 | C |
| *In(3R)Mo* | *3R* | 16563347 | C |
| *In(3R)Mo* | *3R* | 16564891 | A |
| *In(3R)Mo* | *3R* | 16565899 | T |
| *In(3R)Mo* | *3R* | 16825891 | A |
| *In(3R)Mo* | *3R* | 16840241 | T |
| *In(3R)Mo* | *3R* | 16877262 | C |
| *In(3R)Mo* | *3R* | 16881477 | A |
| *In(3R)Mo* | *3R* | 16882614 | C |
| *In(3R)Mo* | *3R* | 16914806 | G |
| *In(3R)Mo* | *3R* | 17081985 | T |
| *In(3R)Mo* | *3R* | 17145087 | G |
| *In(3R)Mo* | *3R* | 17161903 | T |
| *In(3R)Mo* | *3R* | 17183342 | T |
| *In(3R)Mo* | *3R* | 17190382 | C |
| *In(3R)Mo* | *3R* | 17203074 | T |
| *In(3R)Mo* | *3R* | 17226102 | G |
| *In(3R)Mo* | *3R* | 17231109 | T |
| *In(3R)Mo* | *3R* | 17252528 | A |
| *In(3R)Mo* | *3R* | 17255885 | A |
| *In(3R)Mo* | *3R* | 17257625 | A |
| *In(3R)Mo* | *3R* | 17261973 | C |
| *In(3R)Mo* | *3R* | 17346744 | A |
| *In(3R)Mo* | *3R* | 17482849 | T |
| *In(3R)Mo* | *3R* | 17492333 | T |
| *In(3R)Mo* | *3R* | 17512751 | T |
| *In(3R)Mo* | *3R* | 17543357 | A |
| *In(3R)Mo* | *3R* | 17570809 | A |
| *In(3R)Mo* | *3R* | 17574820 | T |
| *In(3R)Mo* | *3R* | 17575776 | T |
| *In(3R)Mo* | *3R* | 17614569 | T |
| *In(3R)Mo* | *3R* | 17618094 | T |
| *In(3R)Mo* | *3R* | 17653963 | A |
| *In(3R)Mo* | *3R* | 17673637 | T |
| *In(3R)Mo* | *3R* | 17731781 | T |
| *In(3R)Mo* | *3R* | 17752308 | T |
| *In(3R)Mo* | *3R* | 17775264 | A |
| *In(3R)Mo* | *3R* | 17798722 | T |
| *In(3R)Mo* | *3R* | 17812150 | A |
| *In(3R)Mo* | *3R* | 17812763 | A |
| *In(3R)Mo* | *3R* | 17833454 | T |
| *In(3R)Mo* | *3R* | 17871386 | A |
| *In(3R)Mo* | 3R | 17878212 | T |
| *In(3R)Mo* | 3R | 17893124 | G |
| *In(3R)Mo* | 3R | 17900659 | A |
| *In(3R)Mo* | 3R | 17905561 | T |
| *In(3R)Mo* | 3R | 17909484 | G |
| *In(3R)Mo* | 3R | 17914642 | A |
| *In(3R)Mo* | 3R | 17915717 | C |
| *In(3R)Mo* | 3R | 18018705 | C |
| *In(3R)Mo* | 3R | 18110219 | T |
| *In(3R)Mo* | 3R | 18151777 | T |
| *In(3R)Mo* | 3R | 18195302 | T |
| *In(3R)Mo* | 3R | 18227258 | T |
| *In(3R)Mo* | 3R | 18229705 | C |
| *In(3R)Mo* | 3R | 18236474 | C |
| *In(3R)Mo* | 3R | 18237459 | A |
| *In(3R)Mo* | 3R | 18248909 | G |
| *In(3R)Mo* | 3R | 18405781 | T |
| *In(3R)Mo* | 3R | 18747568 | T |
| *In(3R)Mo* | 3R | 18755175 | G |
| *In(3R)Mo* | 3R | 19051282 | T |
| *In(3R)Mo* | 3R | 19310873 | A |
| *In(3R)Mo* | 3R | 19540597 | C |
| *In(3R)Mo* | 3R | 19573177 | T |
| *In(3R)Mo* | 3R | 19604547 | T |
| *In(3R)Mo* | 3R | 19614762 | A |
| *In(3R)Mo* | 3R | 19616872 | T |
| *In(3R)Mo* | 3R | 19619722 | G |
| *In(3R)Mo* | 3R | 19621728 | A |
| *In(3R)Mo* | 3R | 19625953 | T |
| *In(3R)Mo* | 3R | 19686653 | A |
| *In(3R)Mo* | 3R | 19690483 | T |
| *In(3R)Mo* | 3R | 19928635 | C |
| *In(3R)Mo* | 3R | 20090826 | G |
| *In(3R)Mo* | 3R | 20102331 | G |
| *In(3R)Mo* | 3R | 20106419 | T |
| *In(3R)Mo* | 3R | 20108509 | G |
| *In(3R)Mo* | 3R | 20712447 | A |
| *In(3R)Mo* | 3R | 20717876 | G |
| *In(3R)Mo* | 3R | 20720722 | C |
| *In(3R)Mo* | 3R | 20761490 | T |
| *In(3R)Mo* | 3R | 20809103 | T |
| *In(3R)Mo* | 3R | 20815949 | C |
| *In(3R)Mo* | 3R | 20837056 | A |
| *In(3R)Mo* | 3R | 21380190 | A |
| *In(3R)Mo* | 3R | 21807559 | A |
| *In(3R)Mo* | 3R | 21956164 | G |
| *In(3R)Mo* | *3R* | 22035252 | T |
| *In(3R)Mo* | *3R* | 22399475 | A |
| *In(3R)Mo* | *3R* | 22436302 | C |
| *In(3R)Mo* | *3R* | 22477725 | G |
| *In(3R)Mo* | *3R* | 22635953 | T |
| *In(3R)Mo* | *3R* | 22660660 | T |
| *In(3R)Mo* | *3R* | 22661217 | A |
| *In(3R)Mo* | *3R* | 22703601 | C |
| *In(3R)Mo* | *3R* | 22850222 | A |
| *In(3R)Mo* | *3R* | 23028130 | G |
| *In(3R)Mo* | *3R* | 23504771 | A |
| *In(3R)Mo* | *3R* | 23589504 | C |
| *In(3R)Mo* | *3R* | 24757430 | T |
| *In(3R)Mo* | *3R* | 24834927 | A |
| *In(3R)Mo* | *3R* | 25052744 | T |
| *In(3R)Mo* | *3R* | 25065632 | T |
| *In(3R)Mo* | *3R* | 25087248 | G |
| *In(3R)Mo* | *3R* | 25206657 | T |
| *In(3R)Mo* | *3R* | 25250616 | A |
| *In(3R)Mo* | *3R* | 25253902 | A |
| *In(3R)Mo* | *3R* | 25293082 | T |
| *In(3R)Mo* | *3R* | 25354278 | T |
| *In(3R)Mo* | *3R* | 25687897 | G |
| *In(3R)Mo* | *3R* | 26584256 | T |
| *In(3R)Mo* | *3R* | 26725477 | A |
| *In(3R)Mo* | *3R* | 26930971 | C |
| *In(3R)Mo* | *3R* | 26933596 | C |
| *In(3R)Mo* | *3R* | 26949382 | A |
| *In(3R)Mo* | *3R* | 26955397 | C |
| *In(3R)Mo* | *3R* | 26960620 | T |
| *In(3R)Mo* | *3R* | 27080067 | G |
| *In(3R)Mo* | *3R* | 27091763 | A |
| *In(3R)Mo* | *3R* | 27114289 | A |
| *In(3R)Mo* | *3R* | 27124527 | T |
| *In(3R)Mo* | *3R* | 27136784 | C |
| *In(3R)Mo* | *3R* | 27266479 | A |
| *In(3R)Mo* | *3R* | 27382123 | C |
| *In(3R)Mo* | *3R* | 27395403 | C |
| *In(3R)Mo* | *3R* | 27395667 | A |
| *In(3R)Mo* | *3R* | 27396540 | T |
| *In(3R)Mo* | *3R* | 27396541 | T |
| *In(3R)Mo* | *3R* | 27419936 | A |
| *In(3R)Mo* | *3R* | 27430813 | G |
| *In(3R)Mo* | *3R* | 27434102 | T |
| *In(3R)Mo* | *3R* | 27434183 | G |
| *In(3R)Mo* | *3R* | 27434363 | C |

Kapun *et al.*, Supporting Information File

| | | | |
|---|---|---|---|
| *In(3R)Mo* | *3R* | 27438021 | T |
| *In(3R)Payne* | *3R* | 12257883 | G |
| *In(3R)Payne* | *3R* | 12259133 | C |
| *In(3R)Payne* | *3R* | 12259894 | A |
| *In(3R)Payne* | *3R* | 12263816 | C |
| *In(3R)Payne* | *3R* | 12289495 | C |
| *In(3R)Payne* | *3R* | 12298324 | A |
| *In(3R)Payne* | *3R* | 12298456 | T |
| *In(3R)Payne* | *3R* | 12316508 | C |
| *In(3R)Payne* | *3R* | 17442150 | T |
| *In(3R)Payne* | *3R* | 20343494 | A |
| *In(3R)Payne* | *3R* | 20562004 | T |
| *In(3R)Payne* | *3R* | 20567442 | G |
| *In(3R)Payne* | *3R* | 20567659 | C |
| *In(3R)Payne* | *3R* | 20567832 | C |
| *In(3R)Payne* | *3R* | 20575824 | G |
| *In(3R)Payne* | *3R* | 20580991 | A |
| *In(3R)Payne* | *3R* | 20580995 | T |
| *In(3R)Payne* | *3R* | 20590675 | G |
| *In(3R)Payne* | *3R* | 20591144 | A |

270 **Supporting Table 4. Inversion frequencies during the experimental evolution experiment.** Inversion frequencies estimated from Pool-Seq

271 data using inversion-specific SNP markers in our laboratory natural selection experiment. Shown are median and average (in parentheses) of

272 allele frequencies for each population.

| Generation | Treatment | Replicate | *In(2L)t* | *In(2R)Ns* | *In(3L)P* | *In(3R)C* | *In(3R)K* | *In(3R)Mo* | *In(3R)P* |
|---|---|---|---|---|---|---|---|---|---|
| 0 | hot | 1 | 0.39 (0.43) | 0.1 (0.11) | 0.16 (0.12) | 0.16 (0.17) | 0.01 (0.01) | 0.04 (0) | 0.21 (0.21) |
| 0 | hot | 2 | 0.39 (0.31) | 0.09 (0.1) | 0.15 (0.25) | 0.15 (0.16) | 0.03 (0.06) | 0.05 (0.08) | 0.19 (0.18) |
| 0 | hot | 3 | 0.43 (0.45) | 0.09 (0.08) | 0.14 (0.13) | 0.15 (0.09) | 0.01 (0) | 0.05 (0.04) | 0.19 (0.17) |
| 15 | hot | 1 | 0.51 (0.56) | 0.05 (0.06) | 0.12 (0.07) | 0.38 (0.57) | 0.02 (0) | 0.08 (0.07) | 0.06 (0.18) |
| 37 | hot | 1 | 0.39 (0.44) | 0.02 (0) | 0.13 (0.2) | 0.41 (0.34) | 0 (0) | 0.06 (0.06) | 0.02 (0.02) |
| 59 | hot | 1 | 0.25 (0.34) | 0 (0) | 0.05 (0.04) | 0.48 (0.5) | 0 (0) | 0.05 (0.05) | 0 (0.04) |
| 15 | hot | 2 | 0.43 (0.44) | 0.03 (0.03) | 0.25 (0.19) | 0.36 (0.25) | 0 (0) | 0.04 (0) | 0.07 (0.02) |
| 37 | hot | 2 | 0.25 (0.12) | 0 (0) | 0.12 (0.31) | 0.36 (0.27) | 0 (0) | 0.03 (0) | 0.02 (0) |
| 59 | hot | 2 | 0.25 (0.28) | 0 (0) | 0.03 (0) | 0.27 (0.24) | 0 (0) | 0.07 (0.05) | 0.01 (0.01) |
| 15 | hot | 3 | 0.52 (0.5) | 0.06 (0.04) | 0.22 (0.22) | 0.29 (0.34) | 0 (0) | 0.17 (0.11) | 0.02 (0.01) |
| 27 | hot | 3 | 0.32 (0.22) | 0.04 (0.03) | 0.16 (0.09) | 0.37 (0.16) | 0 (0) | 0.12 (0.11) | 0.02 (0.06) |
| 37 | hot | 3 | 0.37 (0.37) | 0.03 (0.02) | 0.01 (0.05) | 0.39 (0.3) | 0 (0) | 0.1 (0.1) | 0.01 (0.09) |
| 59 | hot | 3 | 0.23 (0.21) | 0 (0) | 0 (0) | 0.5 (0.61) | 0 (0) | 0.01 (0) | 0 (0.02) |
| 15 | cold | 1 | 0.21 (0.21) | 0.01 (0.01) | 0.11 (0.08) | 0.05 (0.08) | 0 (0) | 0.21 (0.16) | 0.19 (0.22) |
| 33 | cold | 1 | 0.39 (0.39) | 0 (0) | 0.07 (0.07) | 0.07 (0.07) | 0 (0) | 0.22 (0.13) | 0.03 (0.03) |
| 15 | cold | 2 | 0.42 (0.42) | 0.06 (0.02) | 0.12 (0.2) | 0.08 (0.06) | 0 (0) | 0.2 (0.18) | 0.07 (0.07) |
| 33 | cold | 2 | 0.21 (0.14) | 0.01 (0.03) | 0.11 (0.12) | 0.16 (0.09) | 0 (0) | 0.24 (0.24) | 0 (0) |
| 15 | cold | 3 | 0.39 (0.39) | 0.09 (0.09) | 0.05 (0.11) | 0.11 (0.03) | 0 (0) | 0.23 (0.28) | 0.06 (0.04) |
| 33 | cold | 3 | 0.56 (0.52) | 0.02 (0) | 0.07 (0.04) | 0.15 (0.15) | 0 (0) | 0.28 (0.35) | 0 (0) |

273 **Supporting Table 5. Inversion frequency differences during experimental**

274 **evolution.** *P*-values from CMH tests performed between the base population and

275 consecutive generations during the experimental evolution experiment. *P*-values were

276 combined by averaging across all marker SNPs for each inversion.

| Inversion | 0_15_hot | 0_37_hot | 0_59_hot | 0_15_cold | 0_33_cold |
|---|---|---|---|---|---|
| *In(2L)t* | 0.3259 | 0.4464 | 0.0739 | 0.3081 | 0.5377 |
| *In(2R)NS* | 0.3757 | 0.1298 | 0.0139 | 0.3150 | 0.0209 |
| *In(3L)P* | 0.4246 | 0.2829 | 0.0032 | 0.3877 | 0.2022 |
| *In(3R)C* | 0.0275 | 0.0129 | 0.0012 | 0.2040 | 0.3445 |
| *In(3R)K* | 0.4080 | 0.4394 | 0.2045 | 0.4543 | 0.1755 |
| *In(3R)Mo* | 0.2035 | 0.4997 | 0.4699 | 0.0232 | 0.0071 |
| *In(3R)Payne* | 0.0048 | 0.0132 | 0.0009 | 0.0639 | 0.0000 |

277 **Supporting Table 6. Inversion frequencies in natural populations.** Inversion frequencies estimated from Pool-Seq data using inversion-
278 specific SNP markers for the Australian (Kolaczkowski *et al*. 2011) and North American (Fabian *et al*. 2012) data. Median and average (in
279 parentheses) of allele frequencies for each population.

|  | *In(2L)t* | *In(2R)Ns* | *In(3L)P* | *In(3R)C* | *In(3R)K* | *In(3R)Mo* | *In(3R)Payne* |
|---|---|---|---|---|---|---|---|
| **Florida** | 0.41 (0.38) | 0.01 (0.01) | 0.09 (0.09) | 0.01 (0) | 0 (0) | 0 (0) | 0.49 (0.54) |
| **Pennsylvania** | 0.23 (0.22) | 0.04 (0.05) | 0.01 (0.05) | 0.01 (0) | 0 (0.01) | 0.08 (0.05) | 0.02 (0.06) |
| **Maine** | 0.2 (0.21) | 0.1 (0.11) | 0 (0.04) | 0.02 (0.02) | 0.06 (0.07) | 0.14 (0.14) | 0.01 (0) |
| **Queensland** | 0.2 (0.38) | 0.05 (0.04) | 0.09 (0.08) | 0 (0) | 0 (0) | 0 (0) | 0.23 (0.13) |
| **Tasmania** | 0 (0) | 0 (0) | 0 (0) | 0 (0) | 0 (0) | 0 (0) | 0.05 (0) |

Kapun *et al.*, Supporting Information File

281  **Supporting Table 7. Inversion frequency differences in natural populations.**

282  *P*-values from Fisher Exact Tests (FET) performed between the lowest-latitude

283  population (Florida and Queensland, respectively) and all other populations in North

284  America (Florida-Pennsylvania: FP; Florida-Maine: FM) and Australia (Queensland-

285  Tasmania: QT) (also see Kolaczkowski *et al*. 2011; Fabian *et al*. 2012). *P*-values were

286  combined by averaging across all marker SNPs for each inversion.

287

| Inversion   | FP     | FM     | QT     |
|-------------|--------|--------|--------|
| *In(2L)t*   | 0.1848 | 0.0220 | 0.4987 |
| *In(2R)Ns*  | 0.2692 | 0.0703 | 0.6332 |
| *In(3L)P*   | 0.1172 | 0.0752 | 0.5460 |
| *In(3R)C*   | 0.2043 | 0.3590 | 0.6584 |
| *In(3R)K*   | 0.2500 | 0.1091 | 1.0000 |
| *In(3R)Mo*  | 0.0853 | 0.0089 | 0.7476 |
| *In(3R)Payne* | 0.0000 | 0.0000 | 0.3516 |

288

**Supporting Table 8. Expected inversion frequency changes due to neutral evolution.** Here, we performed 100,000 simulations of inversion frequency changes as expected due to genetic drift based on a Wright-Fisher model and tested whether the changes were in the expected direction (sign of frequency change) and stronger than observed in the real data. The empirical *P*-value corresponds to the proportion of simulations resulting in stronger inversion frequency changes consistent across all replicates than observed in the real data from the laboratory natural selection experiment. Note that the frequency increases of *In(3R)C* in the hot and *In(3R)Mo* in the cold temperature treatment were significantly higher than expected due to genetic drift (*P*-value < 0.0042; Bonferroni corrected α of 0.05). Additionally, the frequency of *In(3R)P* significantly decreased stronger than expected due to neutral evolution in the cold temperature treatment. All significant results are indicated by an asterisk.

| Inversion | Treatment | Generations simulated | Sign of frequency change | Empirical *P*-value |
|---|---|---|---|---|
| *In(2L)t* | cold | 33 | - | 0.2105 |
| *In(2L)t* | hot | 59 | - | 0.0302 |
| *In(2R)NS* | cold | 33 | - | 0.0577 |
| *In(2R)NS* | hot | 59 | - | 0.1352 |
| *In(3L)P* | cold | 33 | - | 0.0994 |
| *In(3L)P* | hot | 59 | - | 0.0821 |
| *In(3R)C* | cold | 33 | - | 0.2033 |
| *In(3R)C* | hot | 59 | + | 0.0031* |
| *In(3R)Mo* | cold | 33 | + | 0.0002* |
| *In(3R)Mo* | hot | 59 | - | 0.5250 |
| *In(3R)P* | cold | 33 | - | 0.0020* |
| *In(3R)P* | hot | 59 | - | 0.0152 |

**Supporting Table 9. Reliability of inversion frequency estimates.** *P*-values of FET tests used to test for significant differences between empirically determined inversion frequencies (via karyotyping) and those estimated from inversion-specific SNP markers. *P*-values were. Note that non of the *P*-values were significant, indicating that the two methods for estimating inversion frequencies did not differ from each other in their reliability.

| Generation | Regime | Rep | *In(2L)t* | *In(2R)Ns* | *In(3L)P* | *In(3R)C* | *In(3R)K* | *In(3R)Mo* | *In(3R)P* |
|---|---|---|---|---|---|---|---|---|---|
| 59 | hot | 1 | 0.29 | 1.00 | 1.00 | 1.00 | 1.00 | 1.00 | 1.00 |
| 59 | hot | 2 | 0.82 | 1.00 | 0.34 | 0.42 | 1.00 | 0.66 | 0.12 |
| 59 | hot | 3 | 0.08 | 1.00 | 1.00 | 0.44 | 1.00 | 1.00 | 1.00 |
| 33 | cold | 1 | 1.00 | 1.00 | 0.72 | 1.00 | 1.00 | 1.00 | 0.14 |
| 33 | cold | 2 | 0.31 | 1.00 | 0.33 | 0.16 | 1.00 | 0.83 | 1.00 |
| 33 | cold | 3 | 0.26 | 0.17 | 0.34 | 0.76 | 1.00 | 0.48 | 1.00 |

Kapun *et al.*, Supporting Information File309 **Supporting Table 10. Allele sharing among karyotypes.** Amount of allele sharing
310 between individuals (numbers 96 and 100) carrying *In(3R)Mo* and individuals with
311 other chromosomal arrangements. We only used SNPs which were polymorphic
312 between individuals 96 and 100 and the other *In(3R)Mo* chromosomes, located in two
313 polymorphic regions within the inversion boundaries; region 1 spanned positions
314 17,300,000 to 19,400,000 and region 2 positions 23,400,000 to 24,200,000.
315

| Chrom. region | Individual | No. of SNPs | *In(3R)C* | *In(3R)Payne* | Standard |
|---|---|---|---|---|---|
| 1 | 96 | 382 | 63.97% | 47.00% | 100.00% |
| 1 | 100 | 1197 | 73.77% | 48.12% | 78.11% |
| 2 | 96 | 374 | 64.97% | 56.15% | 100.00% |

316    **Supporting Table 11. Statistical power of inversion-specific marker alleles in**

317    **terms of estimating inversion frequencies.** Exact *P*-values obtained by sampling

318    from a $\chi^2$-distribution calculated from randomly drawn SNPs by means of CMH tests.

319    Significant *P*-values ($P < 0.05$) indicate that inversion-specific markers performed

320    better than SNPs randomly drawn from within the inversion body.

| Inversion  | 0_15_hot | 0_37_hot | 0_59_hot | 0_15_cold | 0_33_cold |
|------------|----------|----------|----------|-----------|-----------|
| *In(2L)t*    | 0.2628   | 0.9400   | 0.0730   | 0.6709    | 0.9997    |
| *In(2R)NS*   | 0.6501   | 0.0812   | 0.0000   | 0.8527    | 0.0000    |
| *In(3L)P*    | 0.9989   | 0.9881   | 0.0003   | 0.5320    | 0.2976    |
| *In(3R)C*    | 0.0000   | 0.0000   | 0.0000   | 0.0802    | 1.0000    |
| *In(3R)K*    | 0.9727   | 0.9775   | 0.9711   | 0.9039    | 0.8684    |
| *In(3R)Mo*   | 0.6842   | 1.0000   | 1.0000   | 0.0000    | 0.0000    |
| *In(3R)Payne*| 0.0000   | 0.0001   | 0.0000   | 0.0089    | 0.0000    |